\documentclass[twocolumn,showpacs,preprintnumbers,amsmath,amssymb,floatfix]{revtex4-1}

\usepackage{graphicx}
\usepackage{dcolumn}
\usepackage{bm}
\usepackage{url}
\usepackage{amsmath}
\usepackage{verbatim}  
\usepackage{relsize}

\newcommand{\pb}{$^{208}\mathrm{Pb}^{82+}$}
\newcommand{\au}{$^{197}\mathrm{Au}^{79+}$}

\newcommand{\nip}{n_\mathrm{IP}}
\newcommand{\kb}{k_\mathrm{b}}
\newcommand{\frev}{f_\mathrm{rev}}

\newcommand{\exy}{\epsilon_{xy}}

\newcommand{\exyi}{\epsilon_{xyi}}
\newcommand{\exyj}{\epsilon_{xyj}}
\newcommand{\exi}{\epsilon_{xi}}
\newcommand{\exI}{\epsilon_{1x}}
\newcommand{\exII}{\epsilon_{2x}}
\newcommand{\eyI}{\epsilon_{1y}}
\newcommand{\eyII}{\epsilon_{2y}}

\newcommand{\eyi}{\epsilon_{yi}}
\newcommand{\el}{\epsilon_l}
\newcommand{\eli}{\epsilon_{li}}

\newcommand{\vI}{\vect{v}_1}
\newcommand{\vII}{\vect{v}_2}
\newcommand{\drm}{\mathrm{d}}

\newcommand{\betxy}{\beta_{xy}}
\newcommand{\betSt}{\beta^*_{xy}}

\newcommand{\tibsxy}{T_{\mathrm{IBS},xy}}
\newcommand{\tibsx}{T_{\mathrm{IBS},x}}
\newcommand{\tibsy}{T_{\mathrm{IBS},y}}
\newcommand{\tibsz}{T_{\mathrm{IBS},z}}
\newcommand{\tibsu}{T_{\mathrm{IBS},u}}
\newcommand{\tibsn}{T_{\mathrm{IBS},N}}
\newcommand{\tibsl}{T_{\mathrm{IBS},l}}
\newcommand{\tradz}{T_{\mathrm{rad},z}}
\newcommand{\tradxy}{T_{\mathrm{rad},xy}}
\newcommand{\tcoll}{T_{\mathrm{coll}}}

\newcommand{\lum}{\mathcal{L}}
\newcommand{\lumsc}{\mathcal{L}_{sc}}

\newcommand{\vect}[1]{\vec{#1}}

\newcommand{\madx}{\textsc{mad-x}}

\begin{document}


\title{Time evolution of the luminosity of colliding heavy-ion beams \\
in BNL Relativistic Heavy Ion Collider and CERN Large Hadron Collider}


\author{R.~Bruce}
 \email{roderik.bruce@cern.ch}
\author{J.M.~Jowett}
\affiliation{CERN, Geneva, Switzerland}
\author{M. Blaskiewicz}
\author{W. Fischer}
\affiliation{BNL, Upton, NY, USA}

\date{\today}

\begin{abstract}

We have studied the time evolution of the heavy ion luminosity and bunch intensities in the Relativistic Heavy Ion Collider (RHIC), at BNL, and in the Large Hadron Collider (LHC), at CERN.  First, we present measurements from a large number of RHIC stores (from Run 7), colliding 100~GeV/nucleon \au\ beams without stochastic cooling. These are compared with two different calculation methods.  The first is a simulation based on multi-particle tracking taking into account collisions, intrabeam scattering, radiation damping, and synchrotron and betatron motion.  In the second, faster, method, a system of ordinary differential equations with terms describing the corresponding effects on emittances and bunch populations is solved numerically. Results of the tracking method agree very well with the RHIC data.  With the faster method, significant discrepancies are found since the losses of particles diffusing out of the RF bucket due to intrabeam scattering are not modeled accurately enough.  Finally, we use both methods to make predictions of the time evolution of the future \pb\ beams in the LHC at injection and collision energy.  For this machine, the two methods agree well.

\end{abstract}


\maketitle

\section{Introduction}

During the design and operation of a heavy-ion collider, e.g., the
Relativistic Heavy Ion Collider (RHIC)~\cite{hahn03} or the Large Hadron
Collider (LHC)~\cite{lhcdesignV1}, one of the main goals is to maximize
the time-integral of the luminosity $\lum$ at the experiments.  As
usual, luminosity is
defined as the event rate per unit cross section and depends only on the
spatial density distributions of the colliding beams at the collision
points.

The time-evolution of the spatial densities is determined via
single-particle dynamics from that of the phase-space distributions,
i.e., the kinetics of the beams.
This, in turn, is determined by the
combined influences of several
inter-dependent physical processes.  
The action of some of these (e.g., beam-beam collisions,
scattering on residual gas) principally remove particles from the beams.
Others (e.g., intrabeam scattering (IBS), radiation damping)
predominantly change their distribution in space and momentum.  To
maximize $\int \lum\,dt$, losses caused by other processes than the
collisions themselves need to be minimized and the beam sizes should
stay small.  These processes therefore need to be understood and modeled
in quantitative detail.


There have  been numerous previous studies of the time
evolution of  luminosity (in colliders) or other performance
parameters
such as the emittances  (e.g., in synchrotron light sources  or the
damping rings of linear colliders).
Among many possible references, we
cite~\cite{hubner85,morsch94,baltz96,epac2004}
which are  particularly close to the applications of this paper.

 Many of these are based on the
solution of systems
of coupled ordinary differential equations (ODEs) that describe the
time evolution   of  a few parameters characterizing the beam
distributions, typically the intensities and the first- and second-order
moments of the distributions (beam centroids and emittances).
Such systems can only be closed with additional assumptions on the
nature of the beam distributions.  Typically these are assumed to be
gaussian in all three degrees of freedom.  Besides closing the system of
equations, this can also allow convenient analytical forms for
some of the terms, e.g., those describing intrabeam scattering or the
way in which the luminosity is modified by crossing angles at the
interaction point.
This approach was  applied to the evolution of heavy-ion luminosity in
the LHC   in Ref.~\cite{epac2004}.
Alternative
approaches involve solving the Fokker-Planck equation for the beam
distribution functions~\cite{wei90,gounder03} and single particle
tracking.  In Ref.~\cite{sidorin06} both the ODE method and particle
tracking were implemented in the same simulation code.

\begin{table*}[tbh!]
\begin{center}
\caption{\label{tab:run7} Typical beam and machine parameters for RHIC Run-7 and the LHC, given for the beginning of store. The transverse emittance in RHIC showed large variations between stores~\cite{zhang07}.}
\begin{tabular}{|l|c|c|c|c|}
\hline
Parameter               & Unit        & RHIC collision & LHC collision & LHC injection \\
\hline \hline
Species                 & ...         & \au & \pb & \pb  \\ \hline
Beam energy             & GeV/nucleon & 100 & 2759 & 177.4 \\ \hline
Lorentz factor $\gamma_\mathrm{rel}$   & ... & 107.4 & 2963.5 & 190.5 \\ \hline
Bunch intensity $N_b$   & $10^9$      & 1.1 & 0.07 & 0.07  \\ \hline
Bunches per beam        & ...         & 103 & 592 & 592  \\ \hline
Normalized transverse  
rms emittance & $\mu$m & 3.1 & 1.5 & 1.4 \\ \hline
Long. rms emittance &  eV s/nucleon & 0.25 & 0.25 & 0.07 \\ \hline  
rms bunch length        & cm          & 30 & 7.94 & 9.97 \\ \hline
rms energy spread       & $10^{-3}$   & 0.8  & 0.11 & 0.39 \\ \hline
No. active interaction points (IPs)& ... & 2--4 & 1--3 & 0 \\ \hline
Optical function at IP $\betxy^*$   & m  & 0.8 & 0.5--0.55 & --- \\ 
Crossing angle at IP    &  $\mu$rad  &  0  & 70--285 & --- \\ 
Peak luminosity         & $10^{27}$ cm$^{-2}$s$^{-1}$ & 3 & 1 & --- \\ \hline
RF harmonic numbers $h$ & ...         & 360, 2520 & 35640 & 35640 \\ \hline
RF gap voltage         & MV          & 3 $(h=2520)$ & 16 & 8   \\
                        &             & 0.3 $(h=360)$ & & \\ \hline
\hline
\end{tabular}
\end{center}
\end{table*}

In this article, we study the time evolution of colliding heavy-ion
beams in
RHIC and LHC.  In Sec.~\ref{sec:measurements}, we present measurements
of luminosity and beam intensities during a large number of physics
stores during Run-7~\cite{drees07} in RHIC and give a summary of the
running conditions.  The data are compared with simulations based on two
different models:
\begin{itemize}

\item Multi-particle tracking:  This method, described in
Sec.~\ref{sec:tracking}, is  rather direct and accurate but slow.
We use
an extended version of the code introduced in
Refs.~\cite{blaskiewicz-cool07,blaskiewicz07,blaskiewicz08}, motivated
by the fact that the longitudinal bunch distributions in RHIC are not
gaussian, as discussed in Sec.~\ref{sec:start-dist}.  It is
important to have a simulation that can model IBS for arbitrary profiles
of bunched beams.

\item ODEs:  Numerical solution of a coupled system of ODEs describing
the beam emittances and bunch populations as in Ref.~\cite{epac2004}.
This method, discussed in Sec.~\ref{sec:diff-eq-model}, is fast but
lacks  accuracy when the beam distributions are not gaussian.

\end{itemize}

In Sec.~\ref{sec:comparison}, both methods are compared with the data
from RHIC, and in Sec.~\ref{sec:lhc} we make predictions for the LHC.

\section{Measured lifetimes in RHIC}
\label{sec:measurements}

The RHIC Run-7 consisted of 191 physics stores in 2007 with colliding \au beams at an energy of 100~GeV/nucleon. The main beam and machine parameters for Run-7 are summarized in Table~\ref{tab:run7}.

During Run-7, longitudinal stochastic cooling became operational for the
Yellow ring~\cite{blaskiewicz08} (the two rings of RHIC   are called
Blue and Yellow).  Stochastic cooling
counteracts the longitudinal diffusion caused by intrabeam scattering
(IBS) that eventually pushes particles outside the
longitudinal acceptance (RF
bucket).  This loss process, discussed in Sec.~\ref{sec:start-dist}, is
responsible for a large fraction of the beam losses in RHIC.  In this
article, we focus on stores without stochastic cooling in order to
make a comparison with the LHC.

\begin{figure}[tbh]
  \centering
\includegraphics[width=6cm]{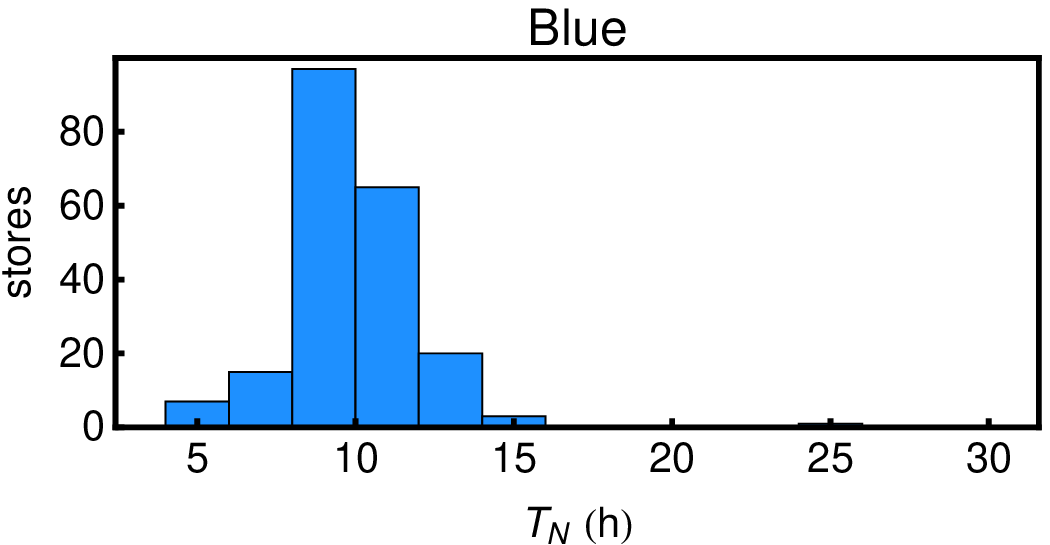}
\includegraphics[width=6cm]{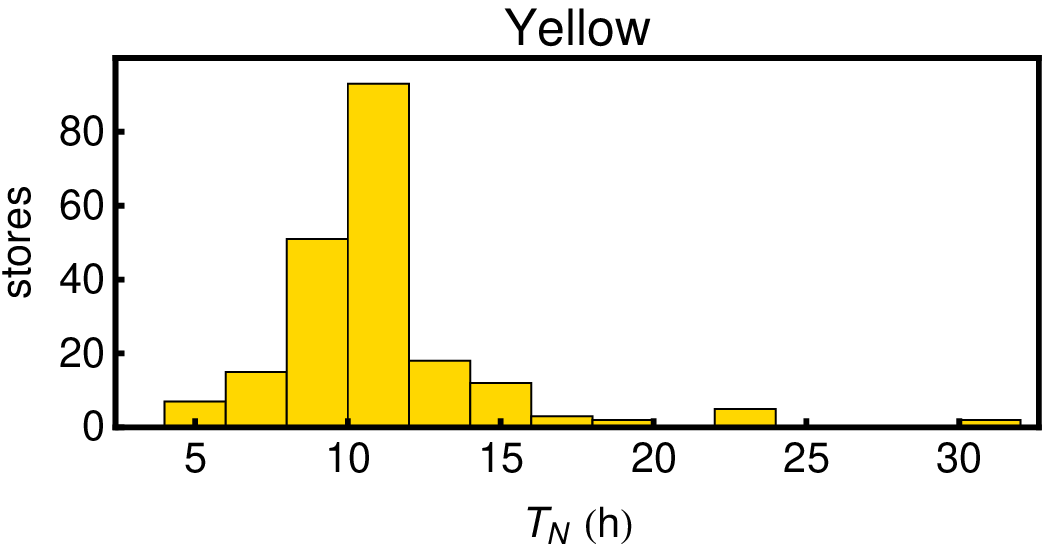}
  \caption{Blue and Yellow beam lifetimes $T_{N}$ obtained by fitting Eq.~(\ref{eq:Exp}) to measurements. Only stores without stochastic cooling are included.
  }\label{fig:blife}
\end{figure}

The time-dependent average bunch intensities $N_i$ ($i=1,2$ for Blue and Yellow respectively) were measured using DC transformers~\cite{hahn03} and can be fitted empirically by an exponential
function
\begin{equation}\label{eq:Exp}
  N_{\mathrm{fit}}(t) = N(0)\exp(-t/T_{N}).
\end{equation}
The beam lifetimes $T_N$ fitted to data from the first 3~h of all physics stores are shown in Fig.~\ref{fig:blife} and the mean and standard deviation values are listed in Table~\ref{tab:lt}. Regular physics stores last 5 h, but some stores terminate prematurely. Almost all stores have data for 3 h. 

\begin{figure}[tbh]
  \centering
  \includegraphics[width=6cm]{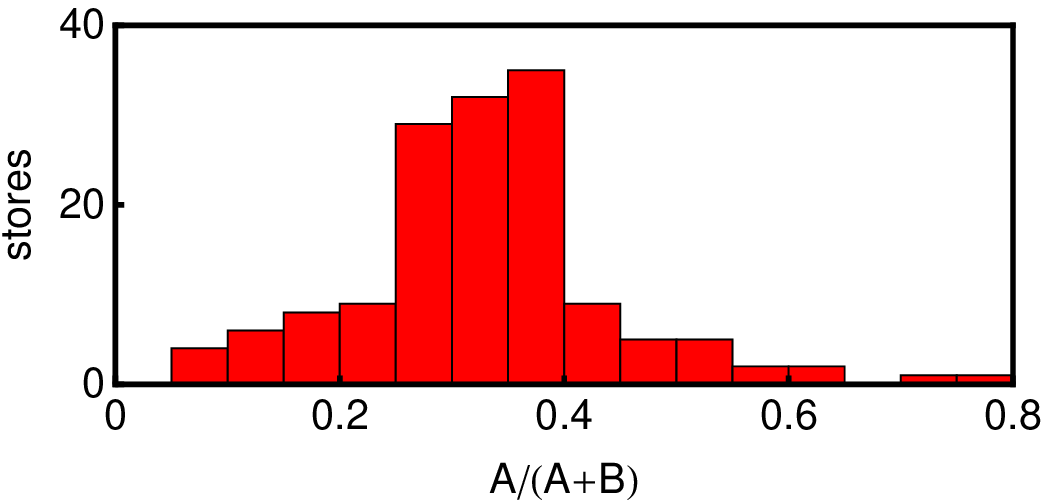}
  \includegraphics[width=6cm]{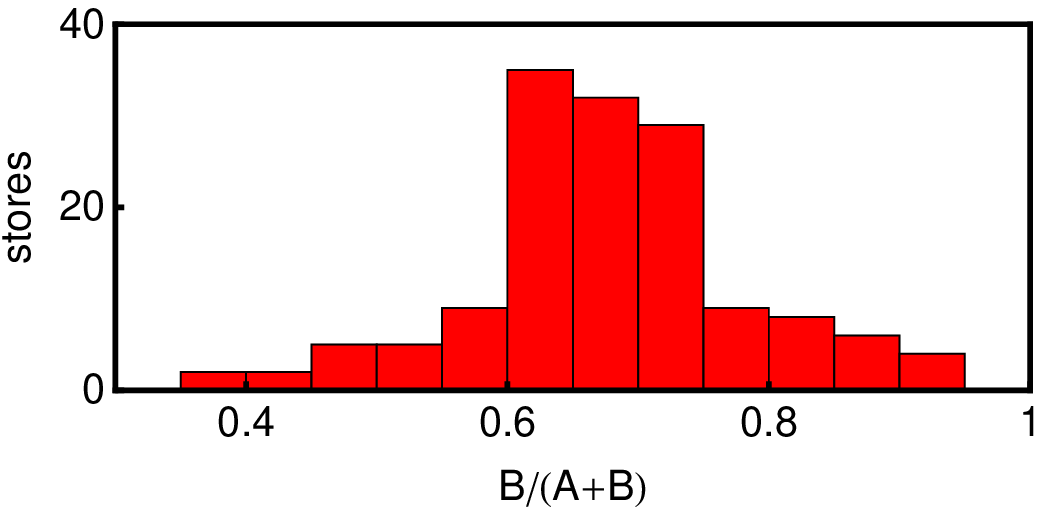}
  \includegraphics[width=6cm]{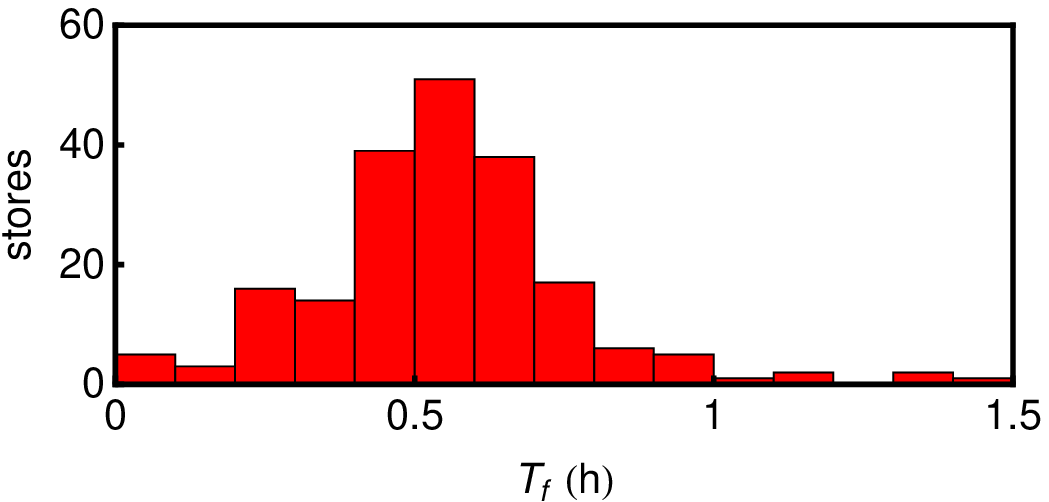}
  \includegraphics[width=6cm]{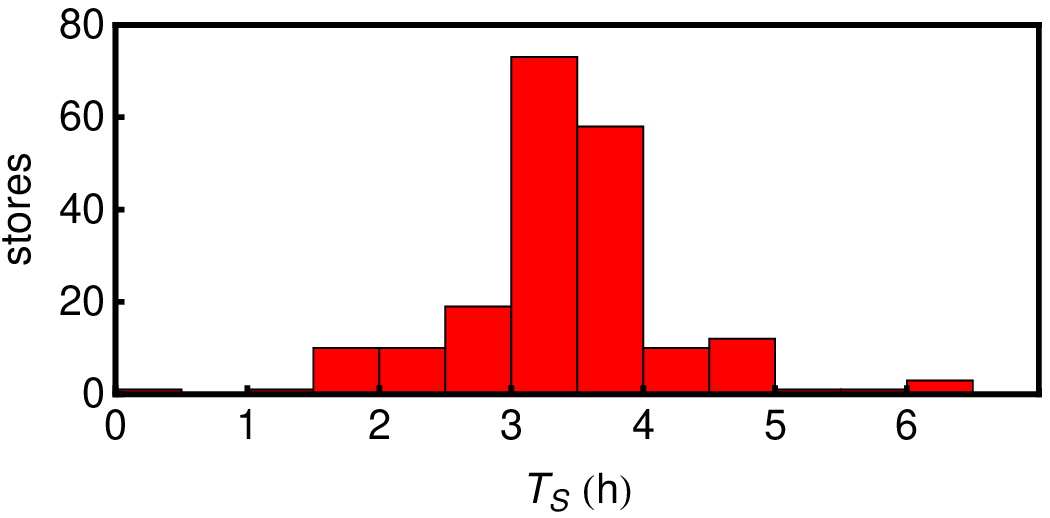}
  \caption{Distributions of fitted luminosity lifetime parameters defined by Eq.~(\ref{eq:doubExp}), where $T_f$ is chosen to be the faster decay time and $T_s$ the slower.
}
  \label{fig:llife}
\end{figure}
\begin{figure}[tbh]
  \centering
  \includegraphics[width=6.5cm]{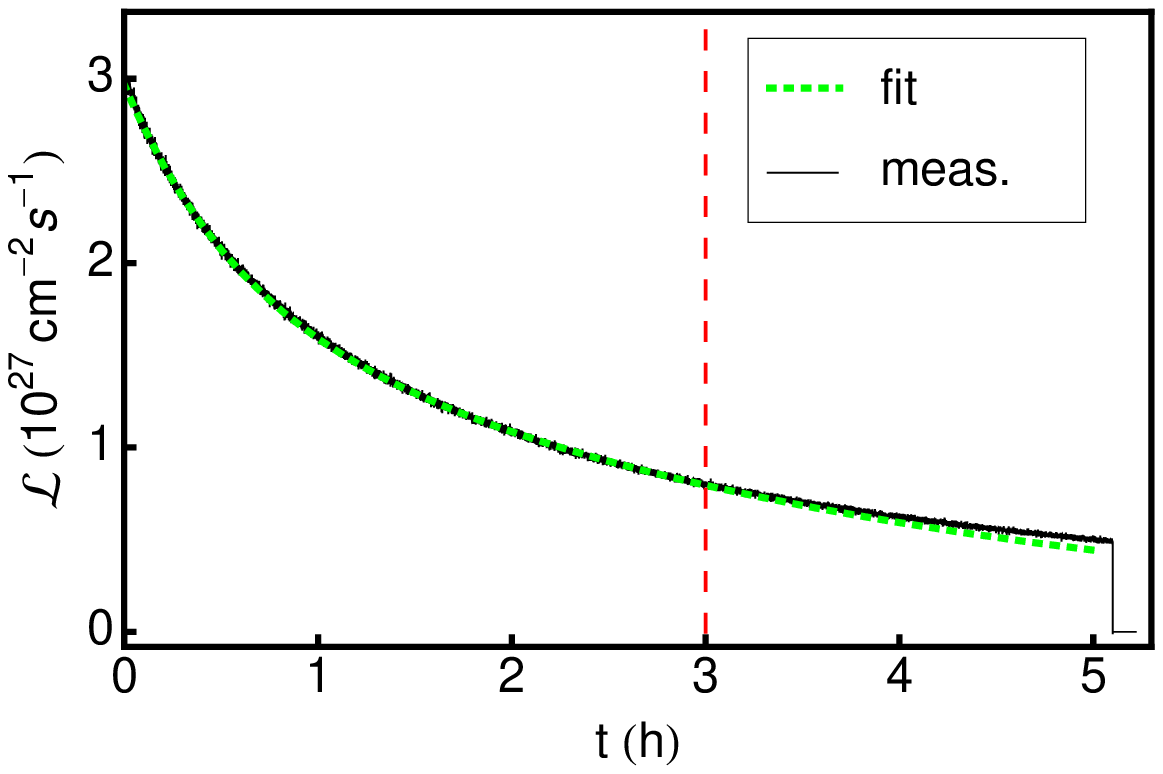}
  \includegraphics[width=6.5cm]{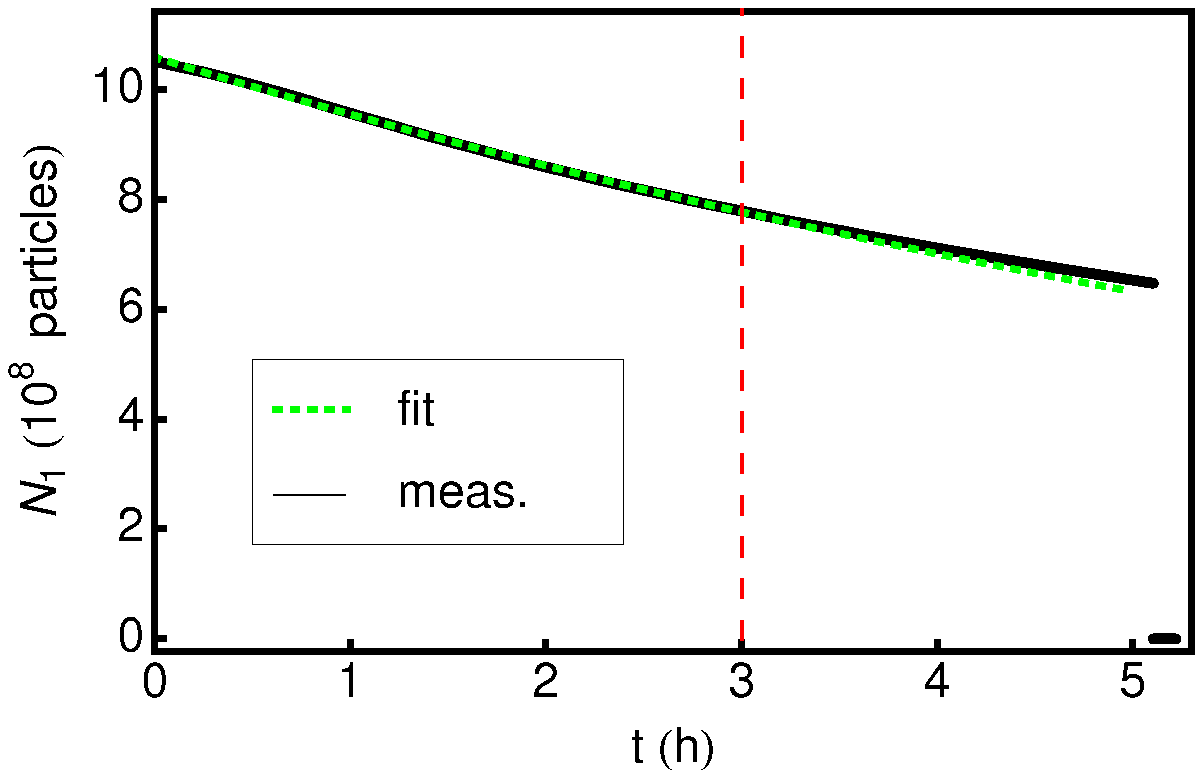}
  \caption{A typical example of measured and fitted luminosity (top) and bunch population (bottom). The fits are done only over the first three hours, indicated by the dashed vertical line, as some stores terminate prematurely. The shown fits have $\chi^2/\mathrm{DOF}=1.2$ and $0.86$. The measurement errors were assumed to be $1.4\times10^{25}$~cm$^{-2}$s$^{-1}$ and $2.5\times10^6$ respectively, given by the spread of measurement points over a short time interval. }
  \label{fig:8908-meas-fit}
\end{figure}

The luminosity was recorded during every store by detecting neutrons
emitted by ions that have undergone mutual electromagnetic dissociation
in the collisions~\cite{baltz98,adler01}.  It turns out that it can be
well fitted by a  sum of fast- and slow-decaying exponential functions
\begin{equation}\label{eq:doubExp}
  \lum_\mathrm{fit}(t)=A \exp(-t/T_f) + B\exp(-t/T_s)
\end{equation}
with the fit parameters $A,T_f,B,T_s$. This is a 3-parameter
fit since $\lum_\mathrm{fit}(0) = A+B$. The fit function (\ref{eq:doubExp}) is purely phenomenological: it is
not chosen on the basis of any particular physical model but rather for the fact that
it generally fits rather well and is convenient to implement.  The fit parameters for all stores are shown in Fig.~\ref{fig:llife} and the means and standard deviations in Table~\ref{tab:lt}.
An example of the measured and fitted luminosity and bunch intensity is shown in Fig.~\ref{fig:8908-meas-fit}.

\begin{table}
 \begin{center}
 \caption{\label{tab:lt} Average and rms values of fitted lifetime
parameters for all 191 physics stores of RHIC Run-7.}
\begin{tabular}{|l|c|c|c|}
\hline
Parameter & Unit & Average & Standard \\
& & & deviation \\ \hline
Blue beam lifetime $T_N$                    & h &  9.9 & 1.7  \\ \hline
Yellow beam lifetime $T_N$                  & h & 10.9 & 3.3  \\ \hline
Luminosity, fast decaying & & & \\
component $A/(A+B)$                         & \%& 33   & 12   \\ \hline
Luminosity lifetime $T_f$, & & & \\
fast decaying                               & h & 0.6  & 0.5  \\ \hline
Luminosity, slow decaying & & & \\
component $B/(A+B)$                         & \%& 67   & 12   \\ \hline
Luminosity lifetime $T_s$, & & & \\
slow decaying                               & h & 3.9 & 2.5   \\ \hline
\hline
\end{tabular}
\end{center}
\end{table}

\section{Particle tracking simulation}
\label{sec:tracking}
The tracking program is based on a code initially used for stochastic cooling~\cite{blaskiewicz-cool07,blaskiewicz07,blaskiewicz08}, which has been developed further to simulate the time evolution of two circulating and colliding bunches. 

Each bunch contains a number of macro particles (in the simulations, $5\times10^4$ were used), which are tracked in a 6D phase space. Normalized coordinates $(x,p_x,y,p_y)$ are used in the transverse planes and $(t,p_t)$ with $p_t=\gamma-\gamma_0$ in the longitudinal ($\gamma_0$ is the Lorentz factor of the reference particle). The particle coordinates are updated on a turn-by-turn basis as a function of the physical processes acting on them. 
The following processes are taken into account:
\begin{itemize}
 \item Synchrotron and betatron motion:  All coordinates are updated deterministically on every turn, taking into account the machine tune, chromaticity and RF voltage. Particles outside the time acceptance of the RF bucket are removed. In RHIC and LHC, these coasting particles are cleaned by extracting them resonantly through a few dipole kicks, sending them onto the collimators. Therefore they have a negligible influence on beam dynamics.
Linear betatron coupling is taken into account in thin-lens approximation. A detailed description can be found in Ref.~\cite{blaskiewicz-cool07}.

\item Collisions:  collision probabilities are sampled as a function of
the local density of the opposing beam at $(x,y)$.  This is described in
Sec.~\ref{sec:track-collisions}.

\item IBS: Each particle is given a random kick, calculated using the Piwinski model, but modulated by the local particle density in order to account for non-gaussian longitudinal bunch profiles. This is described in Sec.~\ref{sec:track-ibs}.

\item Radiation damping and quantum excitation: All particles receive a deterministic amplitude decay and a random excitation. This is described in Sec.~\ref{sec:rad-damp}.

\end{itemize}
Other processes are neglected. 
Beam loss rates in RHIC do not change visibly when the beams are 
brought into collision, and the beam-beam parameter is even smaller 
for ions in the LHC~\cite{lhcdesignV1}. Therefore, the beam-beam effect is 
neglected in our simulations.
A detailed treatment of beam-gas scattering has been done elsewhere for RHIC~\cite{trbojevic97,trbojevic01} and LHC~\cite{jowett04-chamonix,jowett05-chamonix}. In both cases, lifetimes of the order of 100 or several hundreds of hours were found. The emittance blowup at RHIC is fractions of a percent per hour~\cite{trbojevic01} and standard formulas~\cite{handbook98} give a similar result for the LHC with predicted vacuum conditions~\cite{jowett04-chamonix,lhcdesignV1}. Therefore,  beam-gas scattering has a negligible impact on the dynamics in RHIC and LHC when compared to collisions or IBS. 

Furthermore, calculations with the \madx~\cite{madx} program show that the Touschek effect (large-angle scattering from the center of the bucket bringing particles outside the energy acceptance) is negligible in both machines, with lifetimes of hundreds or thousands of hours. However, small-angle scattering of particles already close to the acceptance is important and included as discussed in Sec.~\ref{sec:track-ibs}. 

The strengths of the processes (collision probabilities, inverse radiation damping times, and kicks from intrabeam scattering and quantum excitation) are scaled up to account for both the smaller number of macro particles and by an additional factor, set by the user, that reduces the computational time, so that the number of turns in the simulation corresponds to a much larger number of turns in the real machine. Typically one simulation turn corresponds to $2\times10^4$ real machine turns in our simulations to achieve a relatively fast execution without loss in precision.

\subsection{Collisions}
\label{sec:track-collisions}

A collision probability $P_1$ is calculated for every particle on every turn, and a random number is sampled to determine if an interaction takes place. In that case, the particle is removed. To calculate $P_1$ we perform an overlap integral of the density of the opposing bunch with a Dirac $\delta$-function that represents the trajectory of a single particle. The details of this are shown in Appendix~\ref{app:coll-prob}. The most general collision routine makes no assumptions on the beam distributions and uses a discrete binning of the particles. The integration is then replaced by a sum over the bins. This routine is slow and a much faster code is obtained with some simplifications.

First we assume gaussian distributions in $x$ and $y$, which is generally a very good approximation (in RHIC, the longitudinal distribution is however non-gaussian as explained in Sec.~\ref{sec:start-dist}). Furthermore, we assume a common luminosity reduction factor instead of computing it for every particle, so that $P_1$ is a function of the transverse coordinates only. In the transverse plane we use action-angle variables $(J_x,\phi_x)$ defined for a particle in bunch~1 by
\begin{eqnarray}
\label{eq:ac-ang}
x_1&=&\sqrt{2 J_x \betSt}\cos \phi_x \nonumber \\
x_1'&=&-\sqrt{\frac{2 J_x }{\betSt}}\left(\sin \phi_x+\alpha_{xy}^* \cos \phi_x\right),
\end{eqnarray}
with the angle variables given at the IP and an analogous definition in $y$. Here $\betSt$ is the value of the optical function at the IP (assumed to be the same in $x$ and $y$) where we define the longitudinal coordinate to be $s=0$ and $x'=dx/ds$. We note that $\alpha_{xy}^*=-\betxy'(0)/2=0$.

Since every simulation turn corresponds to a large number of machine turns, where $\phi$ has a uniform distribution on the interval $[0,2\pi]$, we average $P_1$ over $\phi$. The resulting collision probability for a particle in bunch~1, derived in Appendix~\ref{app:coll-prob}, is a function of the betatron actions $J$ only:
\begin{equation}
 \label{eq:coll-phase-avg}
P_1=\sigma N_2 \frac{  \exp\left(-\frac{J_x}{2\exII}-\frac{J_y}{2\eyII}\right)  I_0\left(\frac{J_x}{2\exII}\right)  I_0\left(\frac{J_y}{2\eyII}\right)}
{2\pi\betSt\sqrt{\exII\eyII}}R_\mathrm{tot}
\end{equation}
Here $\sigma$ is the interaction cross section, $N_2$ the intensity of bunch~2, $I_0$ a modified Bessel function, ($\exII,\eyII$) are the geometric rms emittances of bunch~2 and $R_\mathrm{tot}$ is the global luminosity reduction factor, which takes into account the hourglass effect and a crossing angle $2\theta$. Writing the longitudinal distributions of the two bunches, normalized to unity, as $\rho_{zi}(z_i)$, with $z_i$ being the distance to the center of bunch $i$ and $i=1,2$, the reduction factor is given by
\begin{multline}
\label{eq:Rtot}
 R_\mathrm{tot}=\mathlarger{\mathlarger{\int }}
\exp\left({{-\frac{2 \betSt  \sin^2\theta}{(\exI+\exII) \left(1+\frac{2 {\betSt}^2[1+\cos(2 \theta)] }{(z_1+z_2)^2} \right)}}}\right) \\
\times \frac{
\rho_{z1}(z_1)\rho_{z2}(z_2)}
{1+\frac{(z_1+z_2)^2}{4{\betSt}^2\cos^2\theta}} \,\drm z_1 \, \drm z_2.
\end{multline}

When collisions occur at several IPs, $P_1$ is summed up over all of them (only the factor $R_\mathrm{tot}/\betSt$ varies). Furthermore, $P_1$ has to be scaled to account for the smaller number of macro particles and the shorter time scale in the simulation.

If the transverse action in bunch~1 is assumed to be distributed as
$\rho_{1u}=\exp(-J_u/\epsilon_{1u})/\epsilon_{1u}$ for $u=x,y$, equivalent to a gaussian distribution in $u_1$ and $p_{u1}$, we calculate the luminosity as
\begin{align}
\label{eq:lum-tot}
 \lum &= \kb \frev N_1 \int \rho_{1x}\,\rho_{1y} \frac{P_1}{\sigma} dJ_x \, dJ_y  \nonumber \\
      &= \frac{\kb \frev N_1 N_2 }{2\pi\betSt\sqrt{(\exI+\exII)(\eyI+\eyII)}}R_\mathrm{tot},
\end{align}
where $\kb$ is the number of bunches circulating at frequency $\frev$ in each ring.
Eq.~(\ref{eq:lum-tot}) agrees with standard formulas~\cite{handbook98}.

\subsection{Interaction cross section}

To calculate $P_1$ we need the total cross section $\sigma$ for
interactions in which colliding particles are lost from the beam.  Apart
from inelastic hadronic interactions, bound-free pair production (BFPP)
and electromagnetic dissociation (EMD) have to be taken into account.
These electromagnetic processes create particles with a charge-to-mass
ratio different from the main beam, which eventually leads to particle
loss.  Detailed discussions of these loss mechanisms in heavy ion
colliders can be found in
Refs.~\cite{baltz96,klein01,pac2003,epac2004,lhcdesignV1,prstabBFPP09}.

Cross sections were calculated for BFPP in Ref.~\cite{meier01} and for
EMD in Ref.~\cite{pshen01}.  We have taken standard values for
the inelastic cross sections as used by the RHIC and LHC experiments~\cite{lhcdesignV1,alice,baltz99,dunlop-private}.
In Table~\ref{tab:cross-sections} we summarize the cross sections of the
different processes that were used in the simulation.  As can be seen,
the majority of the luminosity is used up by BFPP and EMD rather than
the
inelastic interactions, which are the usual object of study of the
experiments.
\begin{table}[tbh]
\begin{center}
\caption{\label{tab:cross-sections} Interaction cross sections $\sigma$ in RHIC and LHC for different processes removing ions from the beam. Values are taken from Refs.~\cite{meier01,pshen01,lhcdesignV1,baltz99}. The EMD cross sections include all decay channels.}
\begin{tabular}{|l|c|c|}
\hline
Process               & $\sigma$ in RHIC (barn) & $\sigma$ in LHC (barn) \\
\hline \hline
BFPP & 117 & 281 \\ \hline
EMD  & 99 & 226   \\ \hline
Inelastic & 7 & 8 \\ \hline \hline
Total  & 223 & 515 \\ \hline

\end{tabular}
\end{center}
\end{table}



%


\subsection{Intrabeam scattering}
\label{sec:track-ibs}

The standard formalisms for describing IBS~\cite{piwinski74,bjorken83,martini85,kubo01,bane02,nagaitsev05} assume gaussian profiles.
To simulate RHIC, where the longitudinal distributions are known to be
non-gaussian, an IBS model has to go beyond this assumption.
Such a  model has been implemented in
Refs.~\cite{blaskiewicz-cool07,blaskiewicz08}, which we summarize here for completeness.
It assumes that the longitudinal and transverse degrees of freedom are
independent and that the transverse distributions remain gaussian.

For simplicity and speed, the IBS routine starts from the Piwinski model~\cite{piwinski74}. Our code gives the user the choice between a smooth lattice approximation, resulting in fast execution, or a slower but more precise calculation taking the full lattice into account. In this case, the optical functions are read in from an external table created by \madx. Furthermore, the IBS rise times can be calculated with the term $\eta^2/\beta_u$ ($\eta$ being the dispersion and $\beta_u$ the optical lattice function) replaced by $\mathcal{H}=[\eta^2+(\beta_u \eta' - \frac{1}{2}\beta_u' \eta)]$ as proposed in Ref.~\cite{bane02}. Here primed quantities represent derivatives with respect to the path length $s$. We call this model the \emph{modified} Piwinski method and the model presented in Ref.~\cite{piwinski74} the \emph{original} Piwinski method.
We define the rise times $\tibsu,u=x,y,z$ by
\begin{equation}
\label{eq:ibs-growth}
 \frac{d\epsilon_u}{dt}=\frac{\epsilon_u}{\tibsu}.
\end{equation}
The rise times are calculated for gaussian distributions  but are then
modulated for every particle by the local beam density
to account for arbitrary longitudinal profiles.
Thus, particles are given normally distributed momentum kicks $\Delta p_u$ in each plane on every simulation turn:
\begin{equation}
 \Delta p_u=r\sigma_{pu}\sqrt{2 \tibsu^{-1} T_\mathrm{rev} \sigma_t \sqrt{\pi} \rho_{t}(t)}
\end{equation}
Here $r$ is a gaussian random number with zero mean and unit standard
deviation, $T_\mathrm{rev}$ the revolution time, $\sigma_t$ the standard
deviation of $t$ and $\sigma_{pu}$ the standard deviation of the
momentum $p_u$ in plane $u$.  It can be shown that if the longitudinal
distribution $\rho_t$, normalized to unity, is gaussian, integrating
over all particles yields exactly the growth given by
Eq.~(\ref{eq:ibs-growth}).  The strength of the IBS kicks are also
scaled up by the time ratio of the simulation.

The rise times at the beginning of a store, calculated with the different IBS models in the tracking code and parameters in Table~\ref{tab:run7}, are shown in Table~\ref{tab:ibs}. We show also results from \madx, as discussed in Sec.~\ref{sec:diff-eq-model} and, for comparison, the rise times calculated using the Bjorken-Mtingwa method~\cite{bjorken83} (with the Coulomb logarithm calculated by \madx). As can be seen, IBS is strong in RHIC and at injection energy in the LHC.

For the cases studied here, the Piwinski models taking the lattice into account agree well with the Bjorken-Mtingwa method, which is considered more general~\cite{bane02}. The smooth lattice approximation, with the average $\beta$-function calculated from the tune, gives significantly shorter rise times, thus overestimating the strength of IBS. The average difference between the modified Piwinski method, taking the lattice into account, and the Bjorken-Mtingwa method is 2.9\% and the maximum difference 5.1\%. Despite some known deficiencies in \madx~\cite{zimmermannPriv}, the code agrees well with the other methods for the cases studied here.

Because of the better agreement with the other more general methods, we use the modified Piwinski method for all simulations unless otherwise stated. We take the found discrepancies as a guide to the maximum uncertainty of the rise times used in the simulation. The influence of the uncertainty of the IBS model on the time evolution of the luminosity and bunch population is very small compared to other error sources. A comparison between two of the models is shown in Appendix~\ref{app:ibs}.

\begin{table}[tbh]
\begin{center}
\caption{\label{tab:ibs} IBS rise times for the emittances in RHIC and LHC, calculated using the original Piwinski model in smooth lattice approximation (Piw. sm.), original Piwinski taking the lattice into account (Piw. latt.), and the modified Piwinski method including the lattice (mod. Piw.). The calculations for RHIC are done for gaussian bunches---these values are modulated for each particle by the local density in the tracking code. Results from the Bjorken-Mtingwa (B-M) method and \madx, used in Sec.~\ref{sec:diff-eq-model}, are also shown. }
\begin{tabular}{|l|r|r|r|}
\hline
                                & RHIC coll.  & LHC coll. & LHC injection \\ \hline \hline
$\tibsx$, mod. Piw. (h)         &  1.86 &   13.8        & 6.15 \\ \hline
$\tibsx$, Piw. latt. (h)        &  2.05 &   14.8        & 6.62 \\ \hline
$\tibsx$, Piw. sm. (h)          &  1.67 &   10.3        & 4.42 \\ \hline
$\tibsx$, \madx   (h)           &  2.06 &   13.2        & 6.77 \\ \hline 
$\tibsx$, B-M (h)               &  1.88 &   14.1        & 5.98 \\ \hline \hline  
$\tibsz$, mod. Piw. (h)         &  1.96 &   8.26        & 2.84 \\ \hline
$\tibsz$, Piw. latt. (h)        &  1.93 &   8.14        & 2.82 \\ \hline
$\tibsz$, Piw. sm. (h)          &  1.50 &   7.52        & 2.50 \\ \hline
$\tibsz$, \madx   (h)           &  2.03 &   7.89        & 3.03 \\ \hline
$\tibsz$, B-M (h)               &  1.99 &   7.84        & 2.78 \\ \hline \hline  
\end{tabular}
\end{center}
\end{table}

\subsection{Radiation damping}
\label{sec:rad-damp}

Radiation damping and quantum excitation are modeled in each plane $u$
by a deterministic decay, given by the emittance damping time
$T_{\mathrm{rad},u}$, and a random excitation, as described in
Ref.~\cite{siemann85}.  Since $T_{\mathrm{rad},u}\gg T_\mathrm{rev}$, we
expand to first order in $T_\mathrm{rev}/T_{\mathrm{rad},u}$, so that
the decay coefficient for one turn becomes
$\exp(-T_\mathrm{rev}/T_{\mathrm{rad},u})\approx
1-T_\mathrm{rev}/T_{\mathrm{rad},u}$.  The quantum excitation is
determined by the usual radiation integrals over the lattice~\cite{handbook98}
and would lead, in the absence of other dissipative effects,  to
stationary gaussian distributions with rms sizes
$\sigma_{\mathrm{eq},u}$.    The corresponding one-turn map for the
momentum coordinate
$p_{t}$ is therefore
\begin{equation}
 p_{t}
\to p_{t}(1-T_\mathrm{rev}/\tradz)+\sigma_{\mathrm{eq},z}r\sqrt{2
\langle r^2 \rangle T_\mathrm{rev}/\tradz}.
\end{equation}
where    $r$ is a unit, zero-mean
random number.  Similar maps are used in the transverse planes.

In both RHIC and LHC, the photons emitted by heavy nuclei are very
soft~\cite{lhcdesignV1} so the $\sigma_{\mathrm{eq},u}$ are several
orders of magnitudes smaller than the real beam and quantum excitation
has no significant effect on the dynamics.  In RHIC, radiation damping
is also negligible but it is important, 
enough to counter
IBS, in the LHC at collision energy~\cite{lhcdesignV1,epac2004}.  The
damping times are summarized in Table~\ref{tab:damping}.

\begin{table}[tbh]
\begin{center}
\caption{\label{tab:damping} Calculated radiation damping times for the emittances in RHIC and LHC. The damping is equal in the transverse planes.}
\begin{tabular}{|l|c|c|c|}
\hline
  & RHIC collision & LHC collision & LHC injection \\ \hline \hline
$\tradz$ (h) & 825 & 6.3 & 23749 \\ \hline
$\tradxy$ (h) & 1650 & 12.6 & 47498 \\ \hline
\end{tabular}
\end{center}
\end{table}

\subsection{Starting distribution}
\label{sec:start-dist}

It is important to choose an appropriate, realistic, starting
distribution of the particles in the tracking; otherwise agreement
between measurement and simulation is poor.  In the transverse plane,
the beams are well approximated by gaussian distributions, both in RHIC
and (it is expected)  LHC, and the initial coordinates are generated by
the code from the
starting emittances.  The data from the RHIC stores does not in fact
include the measured  transverse beam sizes so these have been
 inferred from the logged initial  bunch populations and luminosity
using Eq.~(\ref{eq:lum-tot}).  A strong coupling between the horizontal
and vertical planes, keeping the beams round on a short time scale, was
assumed. This has been observed empirically in RHIC~\cite{blaskiewicz-cool07}, while for the future ion operation in the LHC, this corresponds to the somewhat idealized situation described by Ref.~\cite{parzen88}.
We also
assumed equal transverse beam sizes in every  bunch of the two beams.
Despite  not
being justified by measurements, this reduces the dimensionality of the
problem to a tractable level; there is, in any case, insufficient data
to attempt a
fit to the small variations in intensity and size of individual bunches.
Indeed we have found that, if different beam sizes are used in
the simulation, the luminosity remains approximately unchanged while the
bunch currents show small variations.

\begin{figure}[tb]
  \centering
  \includegraphics[width=6.7cm]{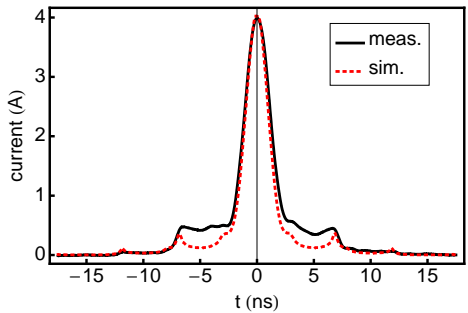}\\
  \caption{Example of a measured bunch profile in RHIC after the ramp when collisions are turned on. We show also the simulated profile used as starting conditions in the simulations. It was generated by letting an initially gaussian bunch circle in the machine under the influence of IBS. 
}
  \label{fig:meas-sim-curr}
\end{figure}

In the LHC, the bunch length is much shorter than the RF bucket size and an approximation of a
gaussian distribution was used. This is not the case in RHIC, as can be seen in Fig.~\ref{fig:meas-sim-curr}, where an example of the bunch current measured at
RHIC when the beams were put into collision is shown.
To understand the bunch shape, we discuss
briefly the synchrotron motion in RHIC.

\begin{figure}[tb]
  \centering
  \includegraphics[width=6.3cm]{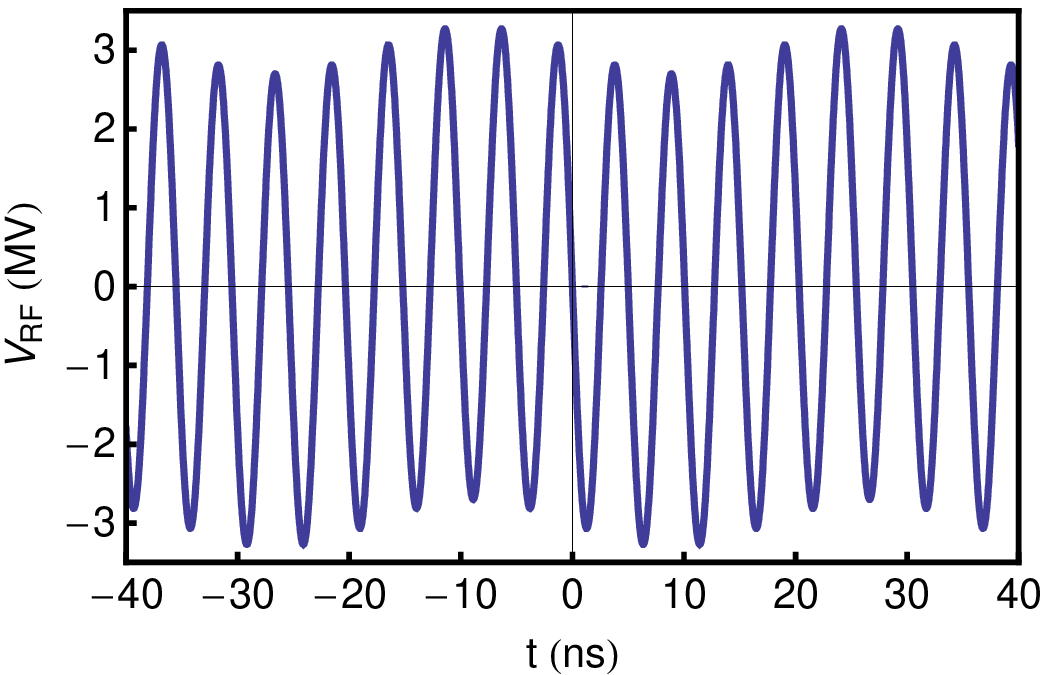}
\includegraphics[width=6.3cm]{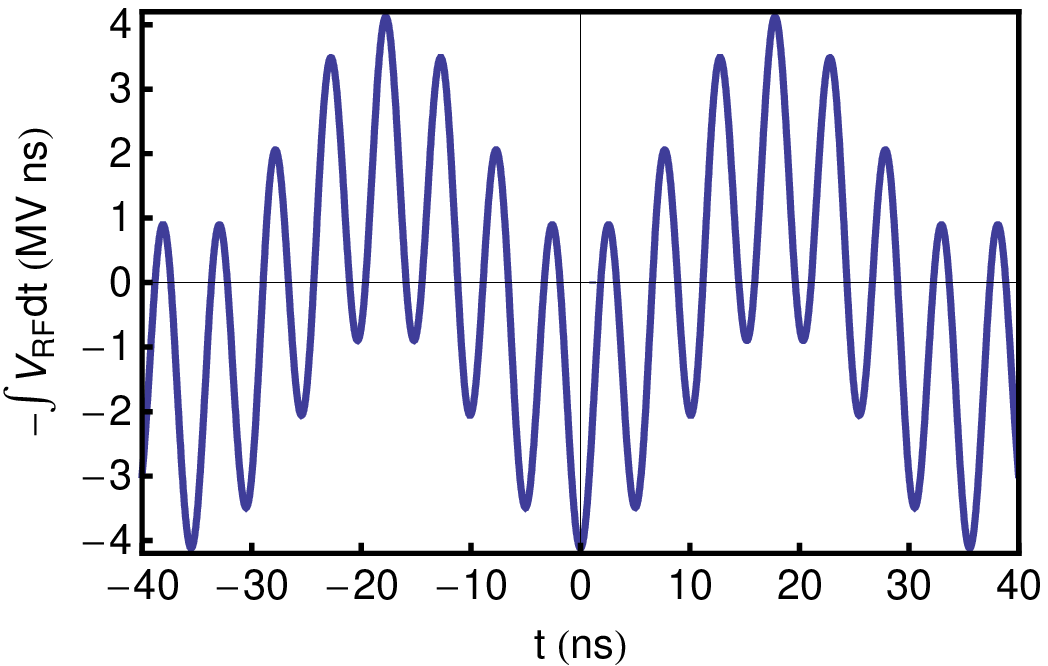}
  \caption{The RF voltage (parameters given in Table\ref{tab:run7}) and its negative time integral. Since synchrotron radiation is negligible, the synchronous phase is at the origin at $V_{RF}\approx0$.}
  \label{fig:rf-volt-potential}
\end{figure}

RHIC uses a double RF system (see Table~\ref{tab:run7}) and the
longitudinal emittance is comparable to the bucket size of the $h=2520$
system.  
The combined RF voltage from both systems is shown in
Fig.~\ref{fig:rf-volt-potential}.  We show also its negative time
integral, which is proportional to the ``potential energy'' term in  the
Hamiltonian of the synchrotron motion.  One can thus picture the
particles ``sliding'' on this surface.  As a particle in the central
bucket continuously receives small momentum kicks from IBS, it
oscillates with higher amplitudes until it has enough energy to enter
the next $h=2520$ bucket.  When it has a high enough amplitude to leave
the $h=360$ bucket, defining the acceptance, the motion becomes
unbounded and it is considered lost by the simulation.  This loss
process is called debunching and is very strong in RHIC.  For this
reason, very significant improvements of the lifetimes and integrated
luminosity  were possible through the implementation of stochastic
cooling~\cite{blaskiewicz-cool07,blaskiewicz07,blaskiewicz08}.

The profile in Fig.~\ref{fig:meas-sim-curr} results from the RF gymnastics performed after the energy ramp, when the storage RF system is turned on. To make the bunches fit in the $h=2520$ buckets, they are shortened through a rotation in the longitudinal phase space, but it is inevitable that some particles escape into neighboring buckets~\cite{blaskiewicz05}.

\begin{figure}[tb]
  \centering
  \includegraphics[width=6.3cm]{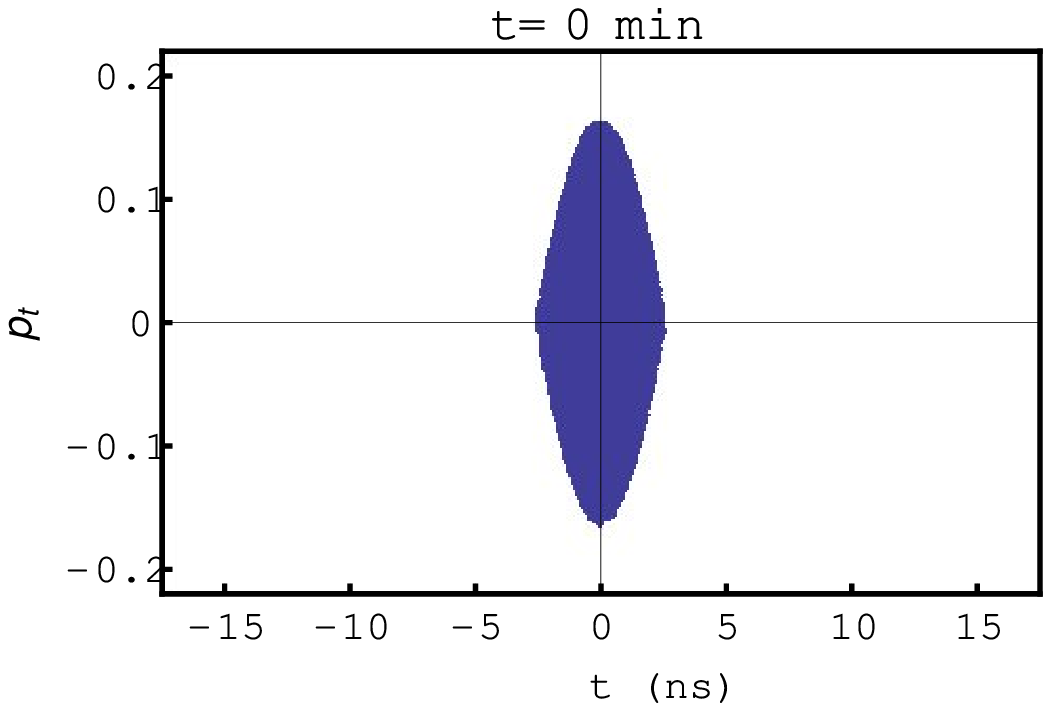}
\includegraphics[width=6.3cm]{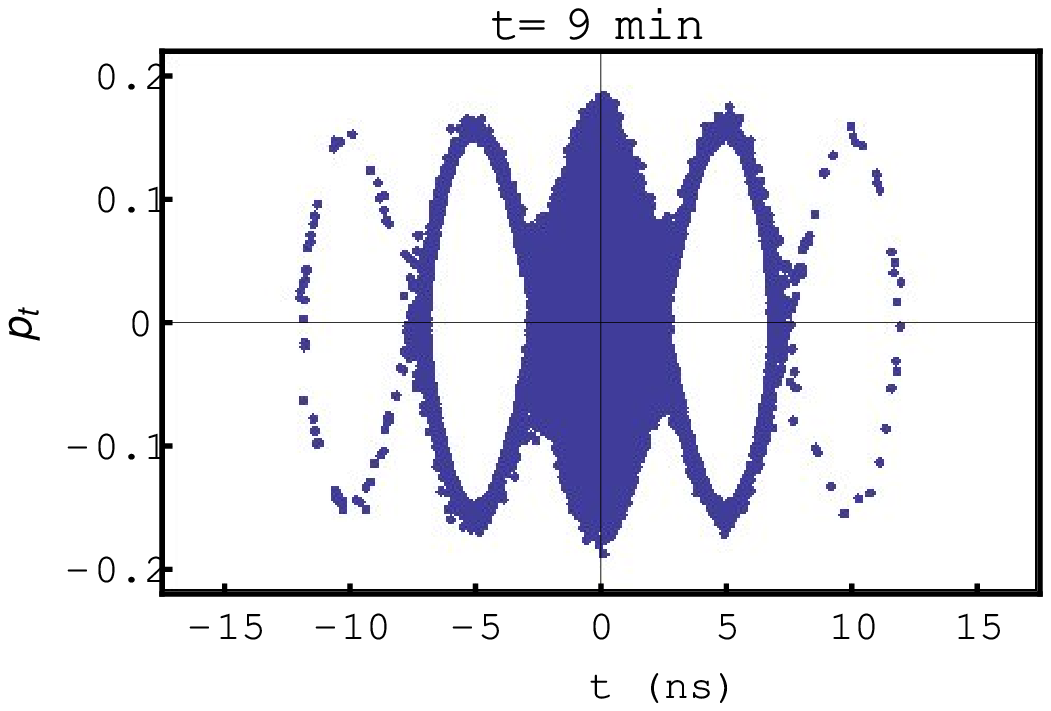}
\includegraphics[width=6.3cm]{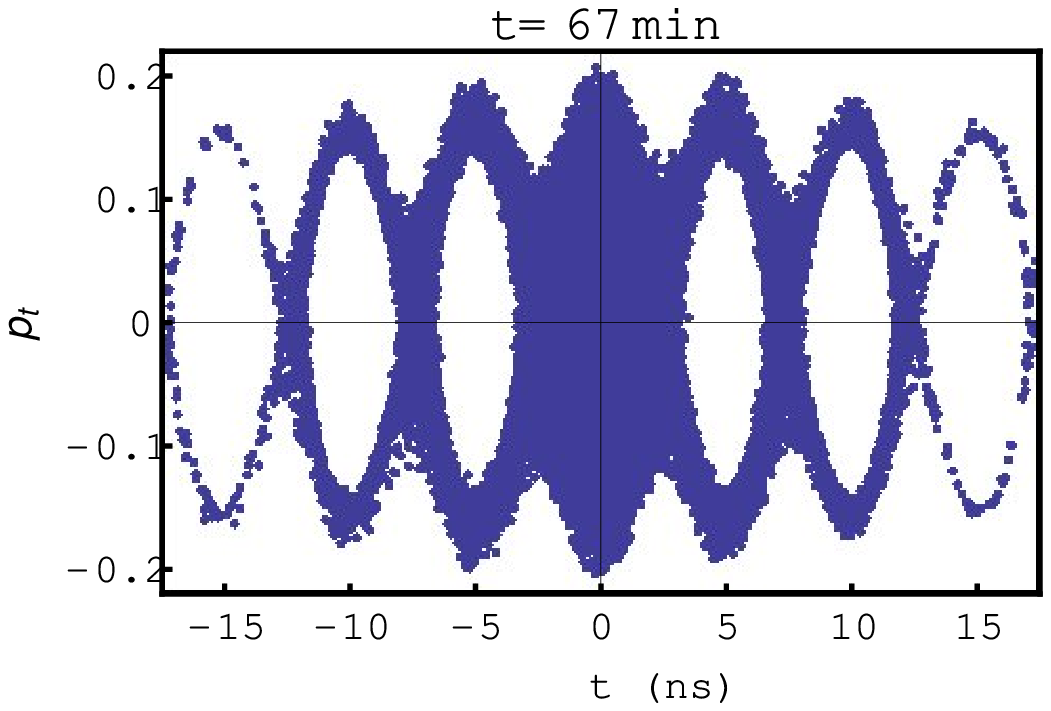}
   \caption{A simulated bunch under influence of IBS only in longitudinal
phase space at different times. Each point represents one particle out
of 50000.  The center of the bunch is located in the central bucket at
$t=0$.  The longitudinal momentum is defined as $p_t=\gamma-\gamma_0$,
where $\gamma_0$ is the Lorentz factor of the synchronous particle.  The
figure shows an idealized situation, where all particles start in the
central bucket (reality is different because of the
RF gymnastics).  The distribution at 67 minutes was used as starting
conditions in the longitudinal plane for the full tracking, due to its
similarity with the measured bunch profile in
Fig.~\ref{fig:meas-sim-curr}. For speed, the smooth lattice approximation was used.}
  \label{fig:long-phase-space}
\end{figure}

To reproduce the profile in Fig.~\ref{fig:meas-sim-curr} and keep the phase space consistent with the RF motion, we used as starting conditions coordinates generated by tracking an initially gaussian bunch, located only in the central $h=2520$ bucket, under the influence of IBS only. In Fig.~\ref{fig:long-phase-space} we show the longitudinal phase space from this simulation on different turns. Only particles with an amplitude high enough to pass from the central $h=2520$ bucket are present in the neighboring buckets. Therefore, the centers of these buckets are empty in phase space. This gives a time profile similar to the expected result of the bunch shortening (see Ref.~\cite{blaskiewicz05}).

Fig.~\ref{fig:meas-sim-curr} shows the resulting bunch current for the population that was used as starting distribution in the RHIC simulations. It corresponds well to the measured profile except around $t=\pm5$~ns, at the centers of the first side buckets (see Fig.~\ref{fig:rf-volt-potential}). The discrepancy means that in the measurement, the phase space distribution is less hollow than in Fig.~\ref{fig:long-phase-space}.


Even though it would be possible to recreate the measured time profile
exactly, this ad-hoc construction would depend on an arbitrary choice of
the energy deviations of the extra particles in the side buckets.  We
prefer to keep the starting conditions as simple as possible in order to
test the predictive power of the code for a large number of different
conditions and therefore use the profile shown in
Fig.~\ref{fig:meas-sim-curr}.

\begin{figure}[tb]
  \centering
  \includegraphics[width=6.7cm]{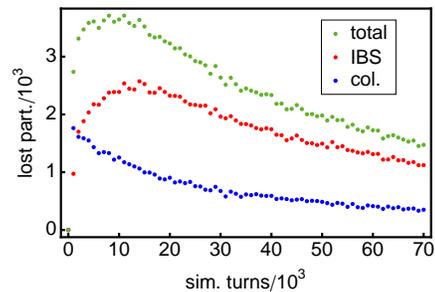}
  \caption{The amount of macro particles lost (integrated over 1000 simulation turns) due to IBS and collisions from bunch 1 (Blue) during the simulation of the test store 8908. In total $5\times10^5$ macro particles were simulated over $7\times10^4$ turns, corresponding to a bunch population of $1.056\times10^9$ and 5.26~h in the real machine.}
  \label{fig:loss-ratio-IBS-coll}
\end{figure}

In Fig.~\ref{fig:loss-ratio-IBS-coll} we show an example of the simulated losses caused by IBS and collisions in RHIC. Simulations show that  if no particles are present in the outer $h=2520$ buckets in the starting population, losses from debunching do not occur from the beginning of the store. With such a longitudinal starting distribution, the initial slope of $N_i(t)$ is less steep and the agreement with measurements is significantly worse.

It should be noted that the assumptions discussed in this section fully determine all free parameters in the simulation.

\section{ODE simulation}\label{sec:diff-eq-model}

As an alternative to the relatively slow tracking simulation we compare
it with a faster method, which does not follow single particles but the
RMS emittances.  This is done by solving numerically a system of coupled
ODEs, similar to Refs.~\cite{kim97,epac2004,sidorin06}.
In all cases treated here, we assume no coherent oscillations of the
bunches                  so that the first-order moments vanish.
Then we assume both
beams to be   gaussian in all degrees of freedom, which
obviously is not true for the longitudinal plane
in RHIC, but  is expected to hold in the LHC.  Solutions for RHIC are
presented in Sec.~\ref{sec:comparison} anyway for comparison.

We assume round beams ($\exi=\eyi\equiv\exyi$ for beams $i=1,2$) due to a strong coupling between the horizontal and vertical planes, as discussed in Sec.~\ref{sec:start-dist}. As in the tracking simulation, we take only collisions, intrabeam scattering and radiation damping into account.  In total we have six dynamical variables: $\exyi,\eli,N_i$, where $\eli$ is the longitudinal emittance. The time evolution of the system is given by:

\begin{align}
\label{eq:ODEs}
\frac{d\exyi}{dt}=&\frac{\exyi}{\tibsxy(N_i,\exyi,\eli)}   -   \frac{\exyi}{\tradxy}   + \frac{\exyi}{\tcoll(N_i,\exyi,\eli)} \nonumber\\  
\frac{d \eli}{dt}=&\frac{\eli}{\tibsl(N_i,\exyi,\eli)}     -   \frac{\eli}{\tradz}\\
\frac{dN_i}{dt}=&-\frac{N_i}{\tibsn(N_i,\exyi,\eli)}-\frac{N_i}{T_\lum(N_i,\exyi,\eli)} \nonumber
\end{align}
%

Here $\tibsxy,\tibsl$ are the instantaneous IBS rise times, and $\tibsn,T_\lum$ the instantaneous lifetimes from debunching and collisions. Furthermore, $\tcoll$ is the emittance rise time caused by core depletion in the collisions~\cite{core-depletion}.  If the transverse distribution is gaussian, the collision probability is much higher for the particles in the core of the beam. When these particles are removed, the remaining ones therefore have a larger transverse emittance. Core depletion is included automatically in the tracking through the calculation of individual collision probabilities for every particle but has to be added explicitly in the ODE system. The core depletion effect is discussed in detail in Ref.~\cite{core-depletion}.  

By adding additional terms to the equations, e.g. as was done for beam-gas scattering in Ref.~\cite{epac2004}, 
 the method can easily be expanded to account for other effects leading to particle loss or emittance growth.

To solve the system numerically or, in certain special cases,
analytically
we use Mathematica~\cite{mathematica}.  Our
implementation allows the kinetics of unequal beam intensities and
emittances to be treated although we shall restrict ourselves to equal
beams in this paper.
When using the method of pre-calculated and interpolated values of
$\tibsl$ and
$\tibsxy$, as explained below,
the gain in execution speed when simulating the time evolution
in a store is more than a factor~1000 compared with the tracking.
A 10~h store takes on the order of a second to solve on a normal desktop computer.



For convenience, we used \madx\ to calculate
$\tibsl$ and $\tibsxy$.  The theoretical framework~\cite{zimmermann05}
is an extended version of the Conte-Martini model~\cite{martini85},  
which includes also vertical dispersion. gaussian beam distributions are
assumed in all planes.  The influence of the IBS model on the simulation result is discussed in Appendix~\ref{app:ibs}, where we make a comparison between \madx\ and the modified Piwinski model.
The evaluation of $\tibsl$ and $\tibsxy$ for
just one set of values $N_i(t),\exyi(t),\eli(t)$ requires a significant
amount of computer time, so it is impractical to do this at runtime. 


We therefore use pre-calculated values of the rise times evaluated on a
grid of points in the relevant region of $(\exy,\el)$.  Since
$T_{\mathrm{IBS}}\varpropto N_i^{-1}$, it is only necessary to evaluate
the rise times for one typical value of $N_i$ and then scale.  The
result is a smooth surface, which can be interpolated with cubic
polynomials in the two emittances.  This initial step, which has to be
done for each optical
configuration and beam energy under consideration, takes less than an
hour of computer time
on a normal desktop machine.  In the Mathematica framework, $\tibsxy$
and $\tibsl$ are replaced by rapidly evaluated interpolating functions
with no loss of precision.  This is the key step that allows a fast
interactive analysis of many cases of interest.  Since we assume round
beams, we use $\tibsxy^{-1}=(\tibsx^{-1}+\tibsy^{-1})/2$ (above
transition in a flat accelerator lattice the uncoupled IBS calculation
gives a large positive growth rate in the horizontal plane and a small
damping rate in the vertical).  The calculated rise times from \madx\
for the initial distributions are shown in Table~\ref{tab:ibs}.



The debunching effect is in principle similar to Touschek scattering,
with the difference that in the standard formulas for the Touschek
effect~\cite{piwinski98} all scattering events leading to a loss are
assumed to occur with large momentum transfers in the center of the
bucket.  The debunching losses are however diffusive in nature, since
most of the lost particles were already very close to the separatrix
before the scattering.  To model this we use a method developed by
Piwinski~\cite{piwinski85}, as follows.


If we assume that the beam distributions change slowly, we can approximate them as being in a steady state during the time step in the numerical integration. The instantaneous lifetime arising when the longitudinal aperture cuts the tails of the gaussian distribution is then, in analogy with Ref.~\cite{piwinski85}, given by
\begin{equation}
\frac{1}{\tibsn}=\frac{\delta_\mathrm{max}^2}{2 \tibsl(N_i,\exyi,\eli) \sigma_\delta^2}\exp\left( -\frac{\delta_\mathrm{max}^2}{2 \sigma_\delta^2}   \right).
\end{equation}
Here $\delta_\mathrm{max}$ is the maximum allowed fractional energy deviation $\Delta E/E_0$ ($E_0$ is the reference energy) and
\begin{equation}
\label{eq:deltaE}
\sigma_\delta^2=\frac{\eli \Omega_S}{\pi \eta_c E_0}
\end{equation}
is the standard deviation of $\Delta E/E_0$, with $\Omega_S$ being the angular synchrotron frequency and $\eta_c$ the frequency slip factor. Eq.~(\ref{eq:deltaE}) is derived for small oscillations, which limits the accuracy of the model. It should be noted that a similar effect exists in the transverse planes due to the cutoff of the physical and dynamic aperture. In the ideal case considered here, without significant impact of resonances and non-linearities, this effect is negligible.


We determine the instantaneous partial lifetime due to the
collisions, $T_\lum$,
from the total number of particles removed from each beam per time:
\begin{equation}
\label{eq:dNdt-lum}
\frac{1}{T_\lum}=-\frac{1}{N_i}\frac{dN_{i}}{dt}=\frac{\lum \sigma}{N_i}
\end{equation}
Here $\lum$ is given by Eq.~(\ref{eq:lum-tot}) and $\sigma$ by Table~\ref{tab:cross-sections}. The reduction factor is calculated from Eq.~(\ref{eq:Rtot}) assuming gaussian distributions in the longitudinal plane. 

Finally, the emittance rise time $\tcoll$ for beam $i$ due to core depletion has been calculated to be~\cite{core-depletion}
\begin{equation}
\label{eq:tcoll}
 \frac{1}{\tcoll}=\frac{\sqrt{\exyi} N_j  \frev \nip R_\mathrm{tot} \sigma}
{4 \sqrt{2} \pi \betSt \left(\exyi+\exyj\right)^{3/2}}.
\end{equation}
Here $i=1,2$ and $j=2$ for $i=1$ and vice versa.

\section{Comparison between simulations and measurements in RHIC}
\label{sec:comparison}

\begin{figure}[tb]
  \centering
  \includegraphics[width=6.3cm]{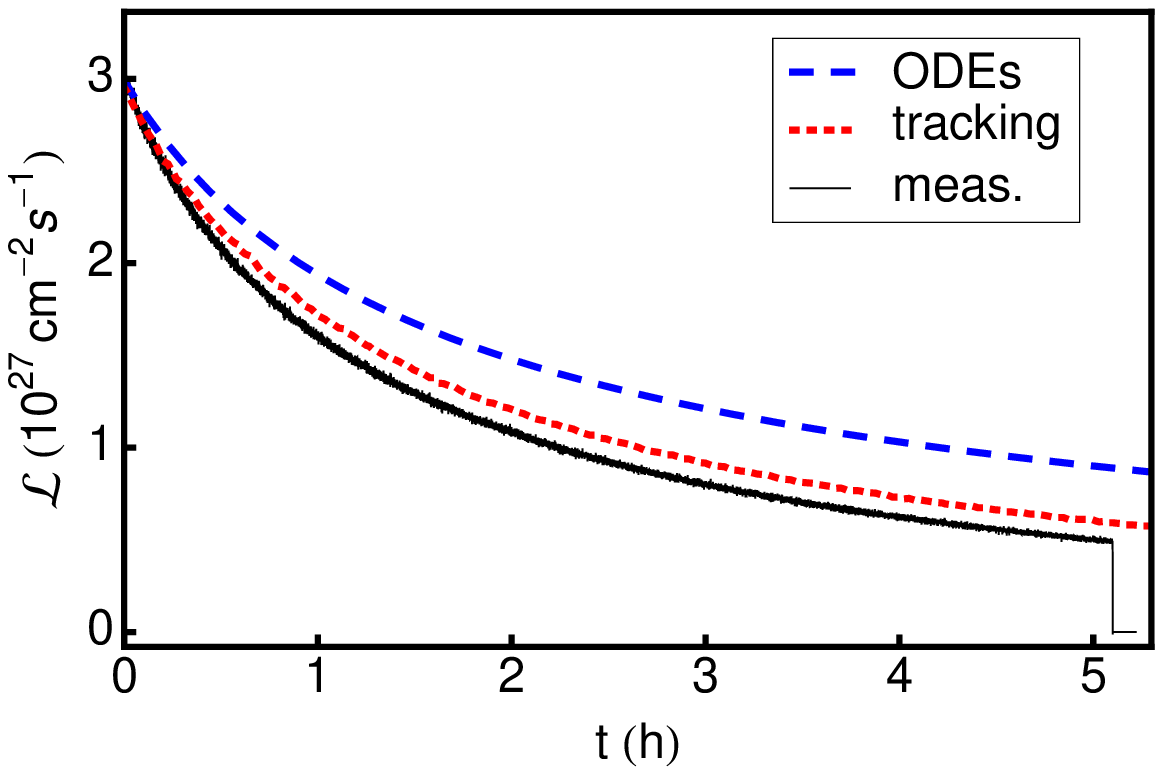}\\
  \includegraphics[width=6.3cm]{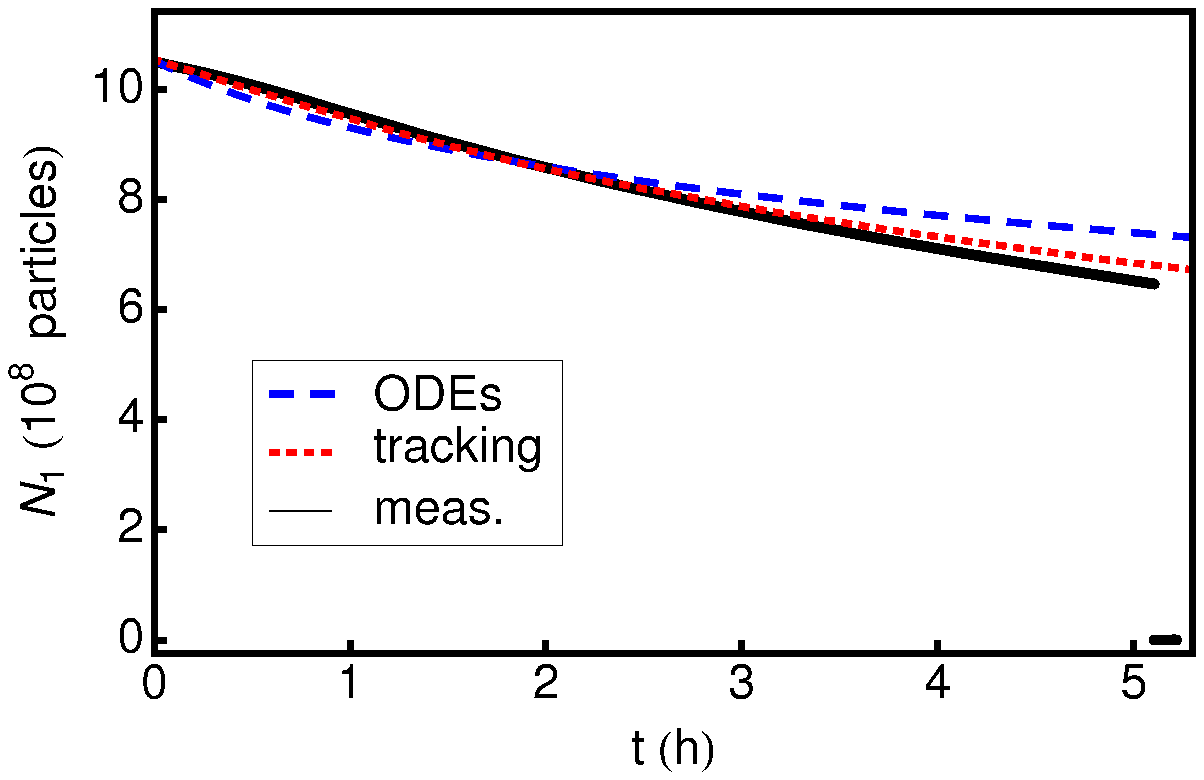}\\
  \includegraphics[width=6.3cm]{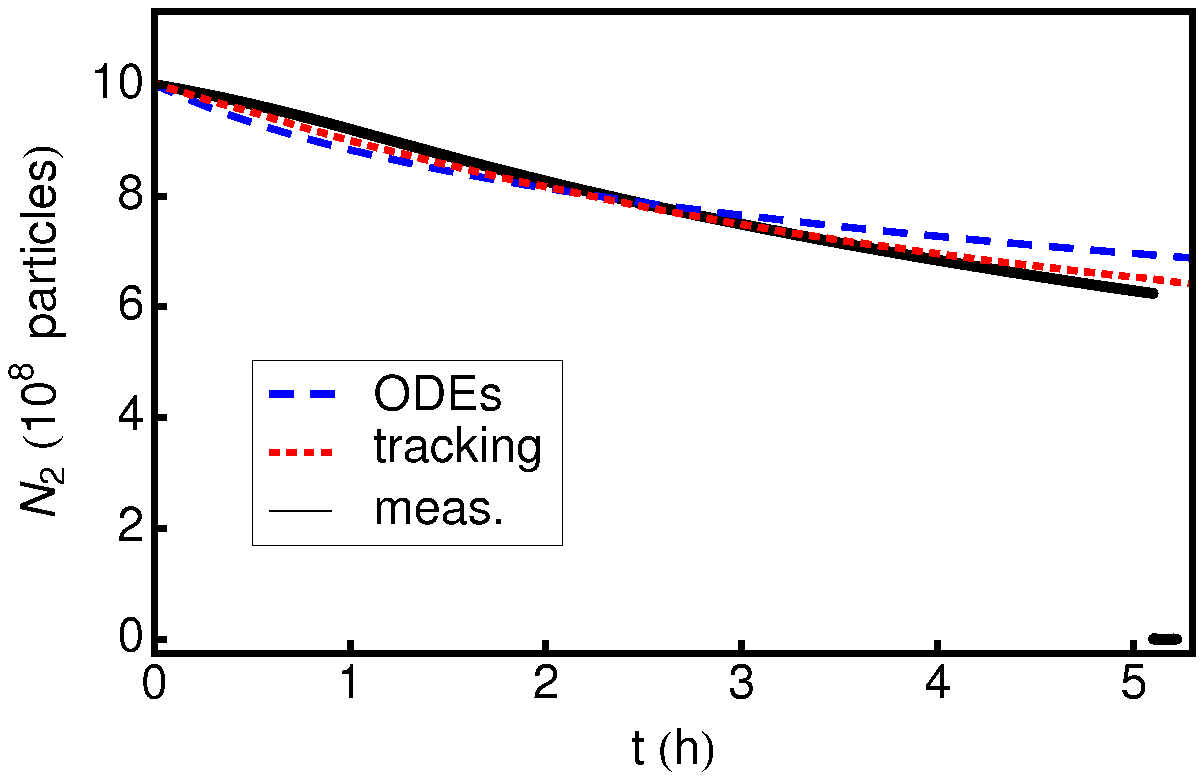}
  \caption{The luminosity (top), Blue bunch intensity (middle) and Yellow bunch intensity (bottom) from measurements and two simulation methods for store 8908 in RHIC. }
  \label{fig:8908-sim-meas}
\end{figure}

As an example of the detailed time evolution in a store, Fig.~\ref{fig:8908-sim-meas} shows
 the measured and simulated luminosity and bunch populations for store 8908 (parameters are given in Table~\ref{tab:8908}). This store was above average in luminosity performance but not exceptionally good. Results from both simulation methods (tracking and ODEs) are shown. A very good agreement is obtained with the tracking method. Using ODEs, the agreement with data is significantly worse. This is clearly because of the limitations in the debunching model and the assumption of gaussian longitudinal profiles. 
In the remainder of this section, we therefore focus on the tracking method only.

\begin{table}[tbh]
\begin{center}
\caption{\label{tab:8908} Starting parameters for the example store called 8908. The transverse starting emittance was not measured, but inferred from the measured luminosity, bunch populations and hourglass factor using Eq.~(\ref{eq:Rh}). It was assumed to be equal for both beams and planes.}
\begin{tabular}{|l|c|c|}
\hline
Parameter & Unit & value \\
\hline
$N_1$ (Blue)        & $10^9$ particles & 1.056\\
$N_2$ (Yellow)      & $10^9$ particles & 1.004\\
$\lum$          & $10^{27}\mathrm{cm}^{-2}\mathrm{s}^{-1}$ & 3.0 \\
transv. geom. rms emittance & mm mrad & 2.34 \\
FWHM of bunch length & ns & 2.6 \\
\hline
\end{tabular}
\end{center}
\end{table}

In order to have more statistics, we simulated, using the tracking method, 139 stores without stochastic cooling with varying bunch populations and luminosity from RHIC Run-7. We use the fits to the measurements, Eqs.~(\ref{eq:Exp}) and~(\ref{eq:doubExp}), for comparisons as they are rather accurate during the first 3~hours. We used the same current, shown in Fig.~\ref{fig:meas-sim-curr}, for all stores, which introduces an error, but allows us to better test the predictive ability of the code.

\begin{figure}[tb]
  \centering
   \includegraphics[width=6.3cm]{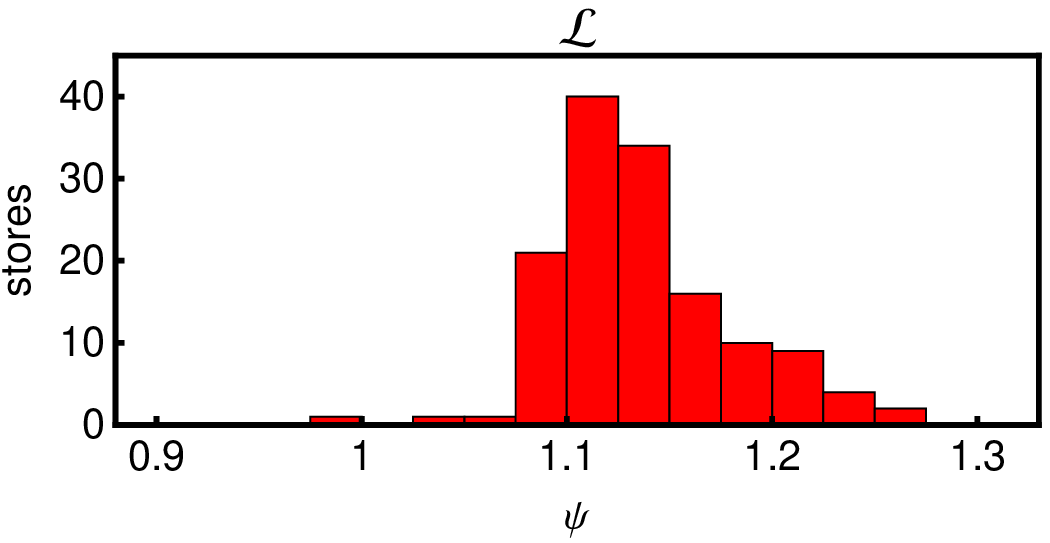}\\
  \includegraphics[width=6.3cm]{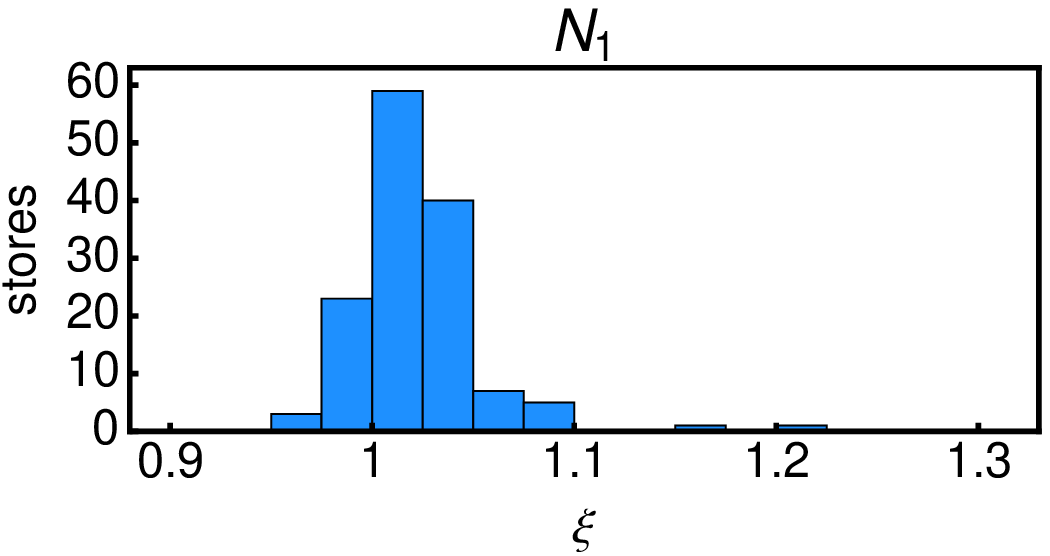}\\
\includegraphics[width=6.3cm]{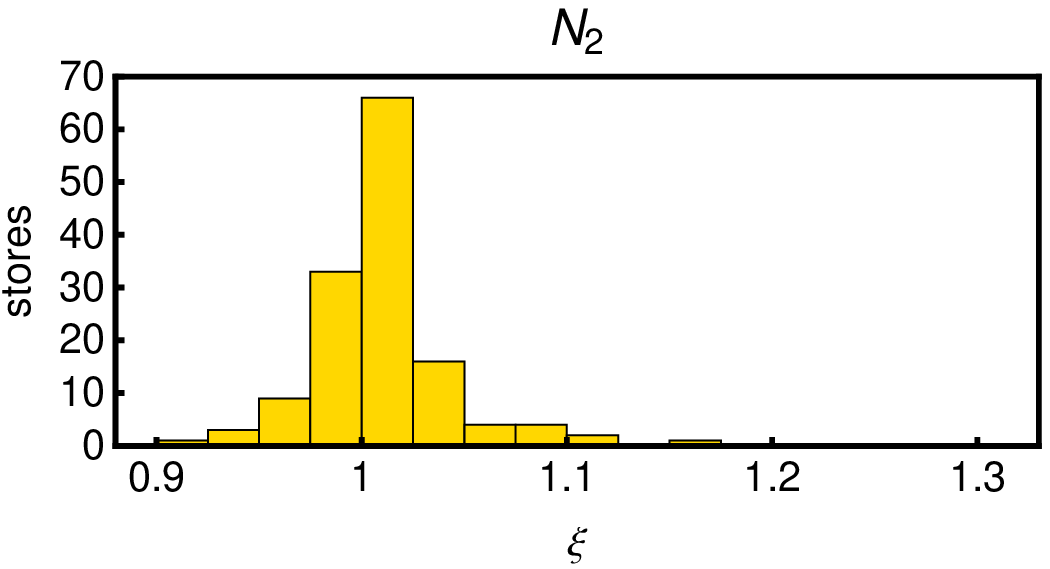}\\
  \caption{Comparison of tracking simulation and measurement of integrated luminosity (top), instantaneous Blue (middle) and Yellow (bottom) bunch intensity for 139~stores in RHIC without stochastic cooling. The parameters $\xi$ and $\psi$ are defined by Eq.~(\ref{eq:xi}).}
  \label{fig:xi-lum-nb}
\end{figure}

To measure the goodness of the tracking simulation, we use two benchmark parameters, $\xi$ and $\psi$, defined as the average instantaneous ratio or ratio of the integrals between a simulated and measured quantity $U$ over the first three hours:
\begin{eqnarray}
\label{eq:xi}
\xi&=&\frac{1}{3\,\mathrm{h}}\int_{t=0\mathrm{h}}^{t=3\mathrm{h}} \frac{U_\mathrm{sim}(t)}{U_\mathrm{fit}(t)} \drm t \nonumber \\
\psi&=&\frac{\int_{t=0\mathrm{h}}^{t=3\mathrm{h}} U_\mathrm{sim}(t) \, \drm t}
         {\int_{t=0\mathrm{h}}^{t=3\mathrm{h}} U_\mathrm{fit}(t) \, \drm t}
\end{eqnarray}
Here $U$ is one of the quantities $\lum,N_1,N_2$. Some histograms of $\xi$ and $\psi$ for 139~stores are shown in Fig.~\ref{fig:xi-lum-nb}. We have deselected 9~stores, where the automatically calculated fit parameters were unrealistic 
or the beam current or luminosity dropped too rapidly to possibly be accounted for by collisions and IBS. 
The mean values and standard deviations of $\xi$ and $\psi$ are shown in Table~\ref{tab:xi-Run7}. An excellent agreement is found for the bunch population, while the integrated luminosity is on average overestimated by 13\%. Given the realities of practical machine operation, we still consider this as a good agreement.

\begin{table}[tbh]
\begin{center}
\caption{\label{tab:xi-Run7} The mean $\langle \xi \rangle$ and the standard deviation $\sigma_\xi$ of the ratio between simulated and measured quantity over 139 stores in Run-7 without stochastic cooling in RHIC.}
\begin{tabular}{|l|c|c|c|c|}
\hline
Quantity & $\langle \xi \rangle$ & $\sigma_\xi$ & $\langle \psi \rangle$ & $\sigma_\psi$ \\ \hline
$\lum$   & 1.16  & 0.056  & 1.13 & 0.044\\
$N_1$    & 1.02 & 0.032  & 1.02 &  0.029\\
$N_2$    & 1.01 & 0.033  & 1.01 &  0.031\\
\hline
\end{tabular}
\end{center}
\end{table}


Several possible sources of the discrepancies between measured and simulated luminosity exist.  Apart from the
uncertainty resulting from the approximations in the simulation model,
and the use of the same longitudinal initial conditions in all stores,
variations of the machine optics, in particular
$\betxy^*$, between different stores could give rise to errors.
The logged data  show that in some
stores, the luminosity varies between the IPs at STAR and PHENIX,
although they have nominally the same $\betxy^*$.


The interaction cross sections have uncertainties of the order of a few 10\%.
Decreasing $\sigma$ to 180~barn in the simulation changes the luminosity
by 4\% after 3~h.
The cross section for mutual EMD, which was used to convert event rate to
measured luminosity, has an uncertainty of not less than about 5\%~\cite{baltz98}.
The uncertainties in cross sections could therefore explain a significant part of
the overestimation of the luminosity in the simulation. 

Calculating IBS rise times in the smooth lattice approximation, which we consider to be a less accurate model that in this case overestimates the influence of IBS, improves agreement with the measured luminosity ($\psi=1.10$) while keeping the excellent agreement with the measured intensities ($\xi\approx1$). This need not mean that the used IBS rise times calculated with the more detailed models are too low. It could also be that some other process, not included in the simulation, causes additional beam losses. Such unaccounted processes include orbit variations, non-linearities, RF noise, ground motion, triplet quadrupole errors, current noise and the effect of dynamic aperture. 
These processes can not easily be included in the simulation, since their strengths are not well known.

\section{Predictions for the LHC}
\label{sec:lhc}

As the overall agreement with data from RHIC is very good, we use the
tracking code to make predictions for the LHC, both at injection and
collision energy.  Run parameters are presented in Table~\ref{tab:run7}.
There are important differences with respect to RHIC.  The LHC uses a
single RF system and bunches are expected to have a gaussian
distribution in the longitudinal plane, meaning that  the ODE method
also has the potential to work well.

\begin{figure}[tb]
  \centering
  \includegraphics[width=6.3cm]{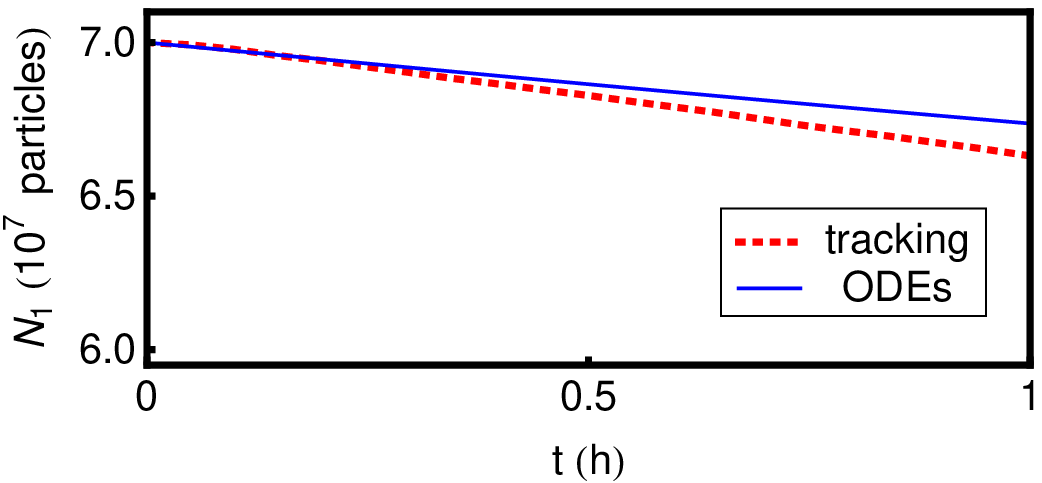}
  \includegraphics[width=6.3cm]{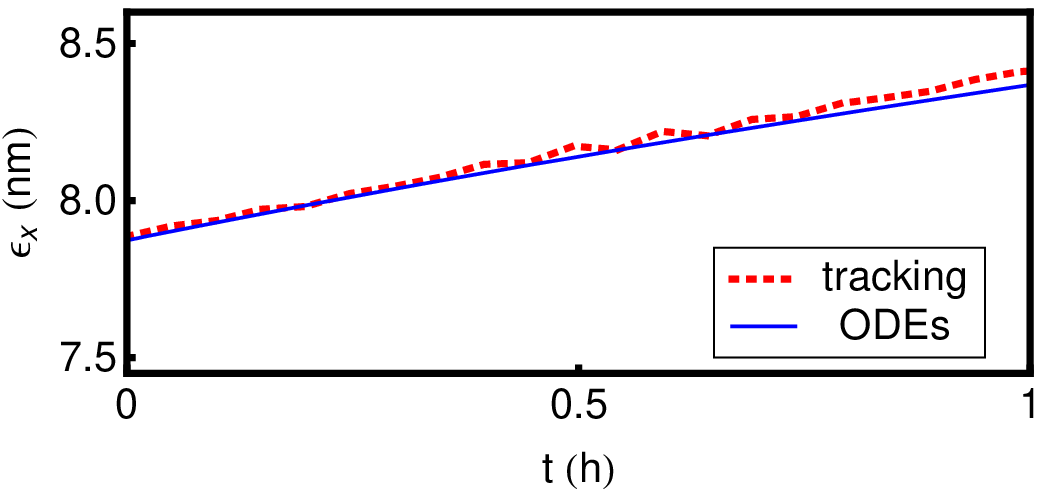}
\includegraphics[width=6.3cm]{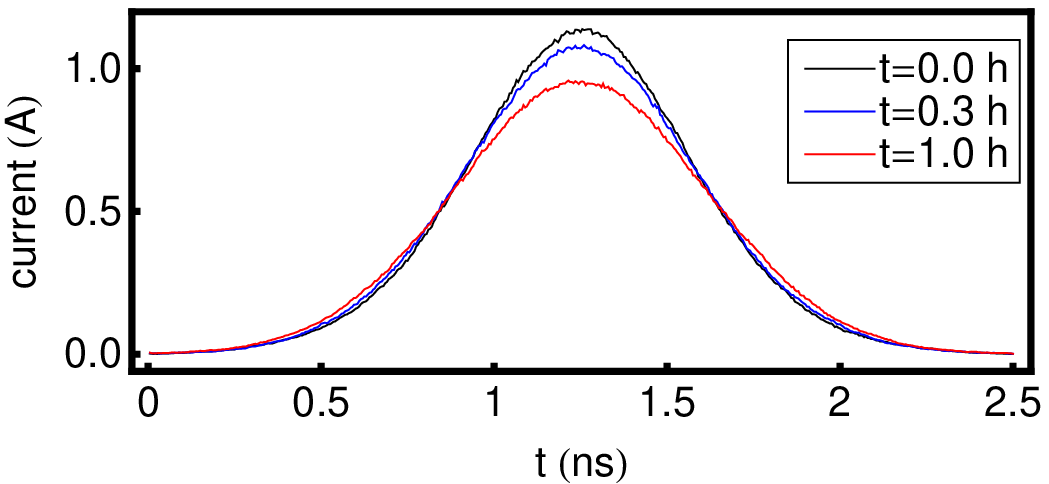}
\caption{The simulated time evolution in the LHC of the bunch intensity and transverse rms emittance during 1~h at the injection plateau without collisions. The bottom plot shows the longitudinal bunch profile at different times as simulated with tracking.}
  \label{fig:lhc-inj}
\end{figure}

In Fig.~\ref{fig:lhc-inj} we show simulations of the time evolution at
the injection plateau, where bunches are kept circulating without
colliding while the ring is filled.  This is expected to take about
10~minutes per ring but, to cover cases where there may be some delay in
starting the ramp, we have simulated
1~h. IBS
rise times are longer than in RHIC, especially in the transverse plane (see Table~\ref{tab:ibs}).  The transverse emittance is
expected to be 2\% larger after 20 minutes and 7\% larger after 1~h.
The simulated bunch current is 1.4\%
lower after 20~min and 5.2\% after
1~h because of debunching losses.  The profile stays approximately
gaussian.

\begin{figure}[tb]
  \centering
  \includegraphics[width=6.3cm]{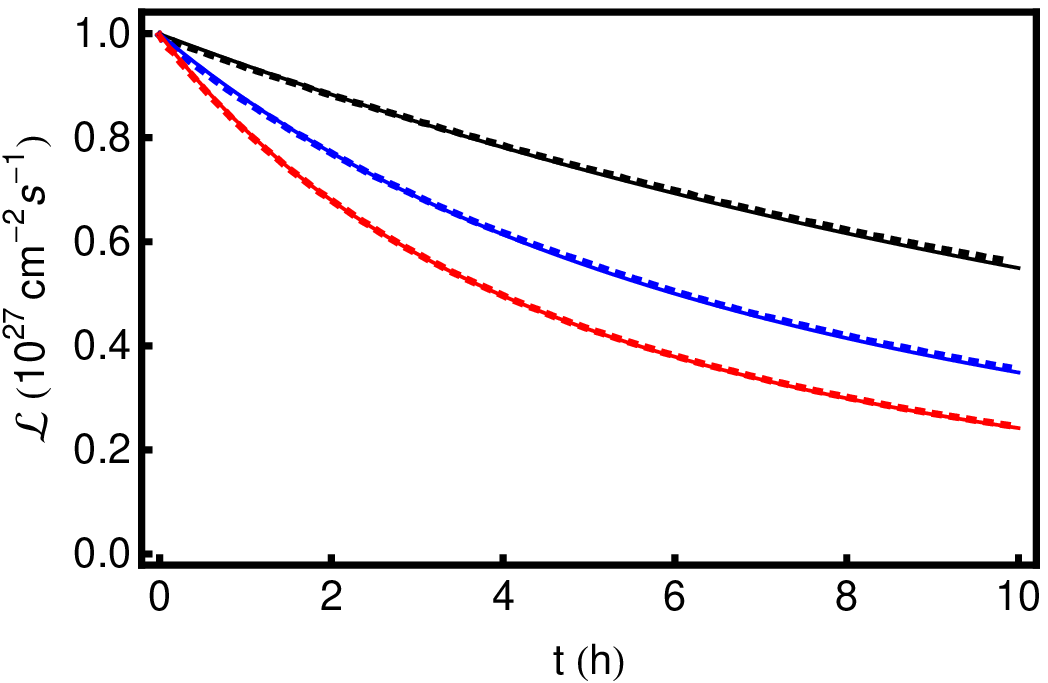}
\includegraphics[width=6.3cm]{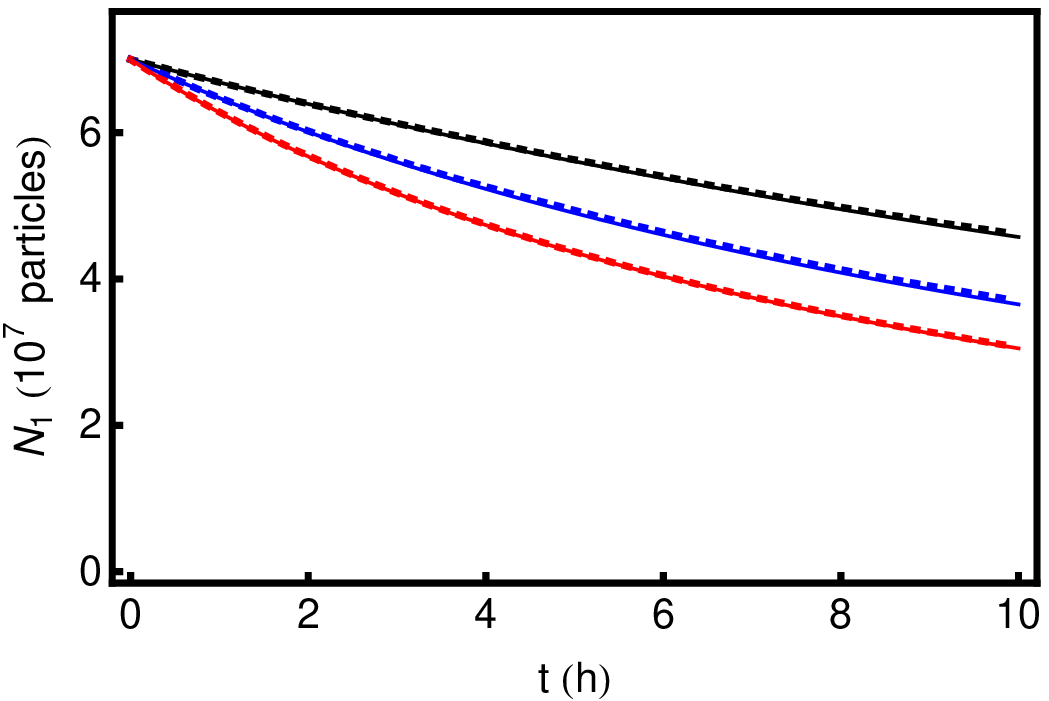}
\includegraphics[width=6.3cm]{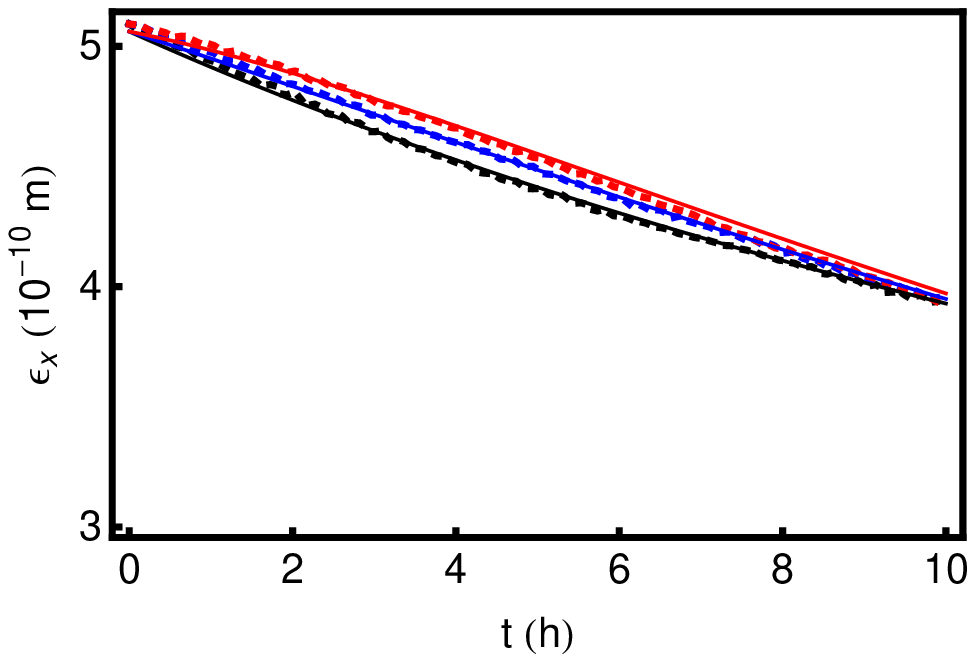}
  \caption{The simulated time evolution in the LHC of the luminosity, bunch intensity and transverse rms emittance during a 10~h store at top energy with colliding beams. Results are shown from the tracking method (dotted lines) and the ODE method (solid lines) for the cases of one (black lines), two (blue lines) or three (red lines) active IPs taking collisions.  }
  \label{fig:lhc-coll}
\end{figure}

Simulations of the time evolution at collision energy using ODEs were
done in Ref.~\cite{epac2004}.  Here we redo the calculations, including also core depletion, and compare
with tracking for 1--3 active IPs (the three experiments ALICE, ATLAS
and CMS are expected to study heavy-ion collisions). The results are
presented
in Fig.~\ref{fig:lhc-coll}.  At collision energy the dynamics in the LHC
changes significantly, since radiation damping (see
Table~\ref{tab:damping}) becomes important and compensates the emittance
blowup from IBS.  As particles are removed from the beams by collisions,
IBS becomes weaker while damping stays constant.  Therefore, the
emittance shrinks during the store.
 The agreement between the simulation methods is
excellent. 
Losses from debunching are predicted to be
negligible in the conditions simulated for the LHC.
Beam losses from collisions are by far the dominant effect.

\section{Conclusions and outlook}
\label{sec:sum}

We have presented measurements of the time evolution of the luminosity
and bunch intensities in both beams during RHIC Run-7 and compared with
two simulation methods:  tracking and ODEs.  
Tracking simulations of 139
stores without stochastic cooling show that the parameter $\xi$, defined
as the average ratio between a simulated and measured quantity, is
close to 1 with a standard deviation of only 3\% for the bunch
populations.  An average over-estimate of the measured integrated luminosity by
around 13\% was found.
Possible sources of the discrepancies include
the use of the same longitudinal starting profile in all stores,
variations in $\betxy^*$, the cross section for 2-neutron
electromagnetic dissociation (used to convert the measured luminosity)
and the neglection of certain factors including RF noise and dynamic aperture.

The ODE simulations show significant discrepancies with measurements in
RHIC, which come from the gaussian approximation of the longitudinal
bunch profile and the limitations in the debunching model.  Including
non-gaussian profiles in this method is not possible in the present
framework.

We have also made predictions of the time evolution in the LHC, where
gaussian bunches are expected and the two simulation methods are in very good
agreement.  In this case, the ODE method increases the execution speed
by a factor 1000.  The dynamics in the LHC is however significantly
different, since radiation damping is a stronger effect than IBS.  This
gives rise to a shrinking emittance at collision energy (although this benefit will vanish
rapidly if the LHC is operated at less than full energy).
Therefore, debunching is not
an issue in the LHC and the collisions are the main source of particle
losses.

Both simulation methods are suitable as testing tools to optimize the
luminosity.  If   stochastic cooling were to be
included, the tracking could be used to make precise predictions of the
improvement in luminosity due to stochastic cooling in one or both beams
in RHIC.

\section{Acknowledgments}
The authors would like to thank J.~Dunlop, A. Fedotov, S. Gilardoni, M.~Giovannozzi, 
P.~Jacobs, A. Sidorin and F.~Zimmermann for helpful discussions.

\appendix
\section{Collision probability}
\label{app:coll-prob}
\begin{figure}[tb]
  \centering
  \includegraphics[width=7cm]{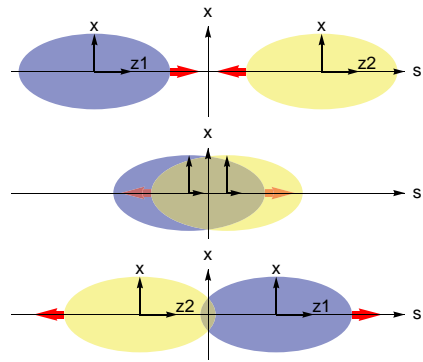}
  \caption{Schematic view of the collision between two bunches at an IP, showing the three coordinate systems used: the $s$-axis is fixed in the laboratory frame and with $s=0$ at the IP, while the $z_1$ and $z_2$ axes move with the origin fixed in the center of the two bunches. The bunch movement is indicated by the red arrows.  The transverse coordinates of all systems coincide, since zero crossing angle is assumed. All distances are measured in the laboratory frame.}
  \label{fig:bunch-crossing}
\end{figure}

To derive the phase averaged collision probability $P_1$ in Eq.~(\ref{eq:coll-phase-avg}) we consider the movement of a particle in bunch~1 through bunch~2 at an IP, using the coordinate systems defined in Fig.~\ref{fig:bunch-crossing} and consider first the case with zero crossing angle. The two bunches move with opposite velocities $v$ so their centers have $s=\pm v\tau$. Both centers are at $s=0$ at the IP at time $\tau=0$. We write the density of bunch~$i$, normalized to one, as $\rho_i(x,y,z_i)$.
The total number of reactions $\lumsc$ per interaction cross section $\sigma$ during a single bunch crossing is~\cite{moller45}:
\begin{multline}
\label{eq:Lsc}
 \lumsc=
M N_1 N_2 \times \\ \int \rho_1(x,y,s-v\tau)\rho_2(x,y,s+v\tau) \,\drm x\,\drm y\,\drm s\,\drm \tau, 
\end{multline}
where $M$ is a kinematic factor defined as
\begin{equation}
 M=\sqrt{(\vI-\vII)^2-(\vI\times\vII)^2/c^2}.
\end{equation}
With zero crossing angle and bunch velocities $|\vI|=|\vII|=v$, we have $M=2v$.

To obtain $P_1$, we define a distribution function $\rho_{p1}$ for a single particle in bunch~1 using the Dirac $\delta$-function. Since zero transverse magnetic field is assumed at the IP, a particle in bunch~1 with spatial coordinates $(x_1,y_1,z_1)$ in the bunch and transverse angles $(x_1',y_1')$ follows a straight line so 
\begin{multline}
 \rho_{p1}(x,y,s-v\tau)= \\ 
\delta(s-v\tau-z_1)\delta(x-[x_1+x_1's])\delta(y-[y_1+y_1's]).
\end{multline}

In the most general collision routine in the tracking code, $\rho_2$ is determined by sorting the particles in bunch~2 in discrete bins along the directions $(x,y,z_2)$. We assume that the transverse distributions at $s=\tau=0$ are independent, writing them as $\rho_{2x}^*(x)\rho_{2y}^*(y)$, and that the longitudinal density $\rho_{2z}$ is independent of $x$ and $y$. The transverse binnings are performed at $s=\tau=0$. 

However, we must take into account that the transverse distributions change along $s$ with the optical function $\betxy(s)$, which we assume to be equal in both planes in the proximity of the IP. In the drift section around the IP $\betxy(s)=\betSt (1+s^2/{\betSt}^2)$, where $\betSt$ is the presumed minimum at the IP. At a given $s$ close to the IP, the transverse distributions are wider by a factor that we call $\kappa(s)$, defined as
\begin{equation}
 \label{eq:kappa}
\kappa(s)=\sqrt{\frac{\betxy(s)}{\betSt}}=\sqrt{1+\frac{s^2}{{\betSt}^2}}.
\end{equation}

Inserting $\rho_{p1}$ in Eq.~(\ref{eq:Lsc}), multiplying by $\sigma$, and integrating, gives $P_1$ for a particle with given coordinates $(x_1,x_1',y_1,y_1',z_1)$:
\begin{equation}
\label{eq:P1-coll1c}
P_1=
2\sigma N_2  \int\frac{\rho_{2x}^*(\frac{x_1+x_1's}{\kappa(s)})}{\kappa(s)}\frac{\rho_{2y}^*(\frac{y_1+y_1's}{\kappa(s)})}{\kappa(s)}\rho_{2z}(2 s-z_1) \,\drm s.
\end{equation}
In the code, the integral in Eq.~(\ref{eq:P1-coll1c}) is replaced by a sum over all bins that a specific particle passes through.

This method is however slow. A faster code is obtained by assuming gaussian distributions in $x$ and $y$. Eq.~(\ref{eq:P1-coll1c}) then becomes
\begin{multline}
\label{eq:P1-Gauss}
 P_1=2 \sigma N_2 \times \\ \int
\frac{\exp\left(-\frac{(x_1+x_1's)^2}{2 \exII \betSt \kappa^2(s)}-\frac{(y_1+y_1's)^2}{2 \eyII \betSt \kappa^2(s)}\right)}
{2\pi\betSt\kappa^2(s)\sqrt{\exII\eyII}}
\rho_{2z}(2 s-z_1) \,\drm s.
\end{multline}
Here the standard deviation of coordinate $u$ in bunch~2 at $s=0$ is given by $\sqrt{\betSt \epsilon_{2u}}$.
We then change to action-angle variables $(J_x,\phi_x)$ using Eq.~(\ref{eq:ac-ang}) and average $P_1$ over $\phi_x$. The $x_1$-dependent exponential in Eq.~(\ref{eq:P1-Gauss}) becomes
\begin{multline}
\label{eq:phase-int}
\mathlarger{\mathlarger{\int}}_0^{2\pi} \exp\left[-\frac{J_x}{\exII}
\left(\frac{\sqrt{\betSt}\cos\phi_x-\frac{s}{\sqrt{\betSt}}\sin\phi_x}
{\sqrt{\betSt+\frac{s^2}{\betSt}}}\right)^2\right]\frac{\drm\phi_x}{2\pi} \\
=\exp\left(-\frac{J_x}{2\exII}\right)I_0\left(\frac{J_x}{2\exII}\right).
\end{multline}
With a similar phase averaging over $\phi_y$, Eq.~(\ref{eq:P1-Gauss}) gives
\begin{multline}
 \label{eq:P1-ph-avg}
P_1=\sigma N_2 \frac{  \exp\left(-\frac{J_x}{2\exII}-\frac{J_y}{2\eyII}\right)  I_0\left(\frac{J_x}{2\exII}\right)  I_0\left(\frac{J_y}{2\eyII}\right)}
{2\pi\betSt\sqrt{\exII\eyII}}\\ 
\times\int \frac{2\rho_{2z}(2s-z_1)}{1+\frac{s^2}{{\betSt}^2}}\drm s.
\end{multline}
Eq.~(\ref{eq:P1-ph-avg}) is the exact phase-averaged collision probability for a particle in bunch~1 with coordinates $(J_x,J_y,z_1)$.  We note that the integral 
represents the hour glass factor for a particle at a given $z_1$.

Furthermore  we approximate the hourglass factor to be equal for all particles, by averaging over all $z_1$-values. Changing integration variable from $s$ to $z_2$ for ease of implementation, the global hourglass factor $R_h$ is
\begin{equation}
\label{eq:Rh}
 R_h=\int \frac{\rho_{z1}(z_1)\rho_{z2}(z_2)}
{1+\frac{(z_1+z_2)^2}{4{\betSt}^2}} \,\drm z_1 \drm z_2.
\end{equation}
It can be shown that, if $\rho_{zu}$ are gaussian, Eq.~(\ref{eq:Rh}) is equivalent to the well-known formulas in Ref.~\cite{furman91}.

With this approximation, $P_1$ becomes
\begin{equation}
\label{eq:P1-coll5a}
 P_1\approx \sigma N_2 \frac{\exp\left({-\frac{J_x}{2\exII}-\frac{J_y}{2\eyII}}\right)I_0\left(\frac{J_x}{2\exII}\right)I_0\left(\frac{J_y}{2\eyII}\right)}
{2\pi\betSt\sqrt{\exII\eyII}} R_h .
\end{equation}

Simulations show that the errors introduced for single particles by using a global $R_h$ are insignificant when considering the whole bunch. For the cases considered in this article, only negligible changes in the simulated time evolution of the luminosity and bunch intensity appear if Eq.~(\ref{eq:P1-coll5a}) is used instead of Eq.~(\ref{eq:P1-coll1c}), while the execution time drops by a factor~15. Therefore, we only use the faster method for the results in this article.

If the bunches collide with a small crossing angle $2\theta$, we replace $R_h$ by a more general luminosity reduction factor $R_\mathrm{tot}$, which accounts for both the crossing angle and the hourglass effect. Such a factor is derived in Ref.~\cite{muratori02} for equal gaussian beams by integration of Eq.~(\ref{eq:Lsc}). Our calculation is completely analogous, with the generalization that we assume unequal transverse beam sizes and unknown longitudinal distributions. The integrations over $(y,\tau)$ can be carried out directly yielding Eq.~(\ref{eq:lum-tot}), with $R_\mathrm{tot}$ given by Eq.~(\ref{eq:Rtot}). 

\section{Comparison of IBS models}
\label{app:ibs}
The simulation results are relatively insensitive to the IBS model. As an illustration, we have calculated the time evolution of the quantities of interest for the LHC collision scheme using the modified Piwinski method for IBS in the ODEs, instead of \madx. Three experiments were assumed active.

\begin{figure}
  \centering
  \includegraphics[width=7cm]{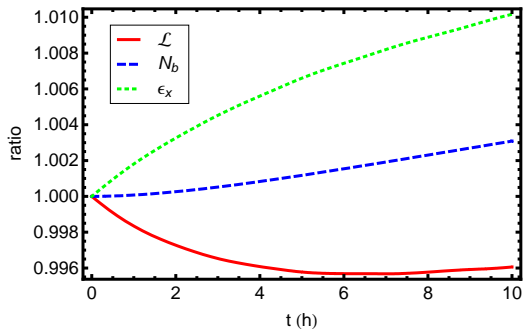}
  \caption{
The ratio of the quantities calculated using \madx\ for IBS to the ones where the modified Piwinski model was used. The time dependence was solved with the ODE method assuming the LHC collision scheme (numerical parameters are shown in Table~\ref{tab:run7}), with three experiments active.}
  \label{fig:lhc-coll-diff-IBS}
\end{figure}

\begin{figure}[tb]
  \centering
  \includegraphics[width=7cm]{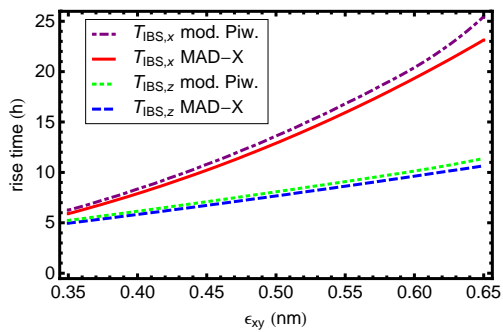}
  \caption{Horizontal and longitudinal IBS rise times for the LHC collision scheme as a function of the transverse emittance as calculated with \madx\ and the modified Piwinski model (mod. Piw.). All other parameters were assumed constant with numerical values shown in Table~\ref{tab:run7}.}
  \label{fig:ibs-times}
\end{figure}

In Fig.~\ref{fig:lhc-coll-diff-IBS}, we show the ratio $U_m/U_p$, where $U$ is the luminosity, bunch population or transverse emittance. The subscript $m$ indicates that \madx\ was used for IBS calculations, while $p$ stands for the modified Piwinski model. The emittance shows the largest deviations (around 1\% after 10~h), which is small compared to the overall error. The IBS rise times as a function of transverse emittance are shown for both models in Fig.~\ref{fig:ibs-times}.

\end{document}